\documentclass[preprint,preprintnumbers,amsmath,amssymb]{revtex4}

\usepackage{axodraw}
\usepackage{graphicx}
\usepackage{dcolumn}
\usepackage{bm}
\newcommand{\be}{\begin{equation}}
\newcommand{\ee}{\end{equation}}
\newcommand{\ba}{\begin{array}}
\newcommand{\ea}{\end{array}}
\newcommand{\bea}{\begin{eqnarray}}
\newcommand{\eea}{\end{eqnarray}}

\newcommand{\lt}{\left}
\newcommand{\rt}{\right}

\newcommand{\chibar}{{\overline\chi}}

\newcommand{\Go}{{\cal G}}

\newcommand{\naively}{na\"\i vely~}
\begin{document}

\preprint{SWAT/04/392}
\preprint{MC-TH-2004-1}

\title{
\phantom{Numerical Portrait of a Relativistic BCS Gapped Superfluid}
\phantom{Numerical Portrait of a Relativistic BCS Gapped Superfluid}
Numerical Portrait of a Relativistic BCS Gapped Superfluid}

\author{Simon Hands}
\email{s.j.hands@swansea.ac.uk}
 \affiliation{Department of Physics, University of Wales Swansea,\\
 Singleton Park, Swansea SA2 8PP, UK}

\author{David N. Walters}%
\email{david@theory.ph.man.ac.uk}
 \affiliation{%
Theoretical Physics Group, Department of Physics and Astronomy,
 University of Manchester, Manchester M13 9PL, UK
}%

\date{13$^{\rm th}$ January 2004
}

\begin{abstract}
We present results of numerical simulations of the 3+1 dimensional
Nambu -- Jona-Lasinio (NJL) model with a non-zero baryon density
enforced via the introduction of a chemical potential $\mu\neq0$. The
triviality of the model with a number of dimensions $d\ge4$ is dealt
with by fitting low energy constants, calculated analytically in the
large number of colors (Hartree) limit, to phenomenological
values. Non-perturbative measurements of local order parameters for
superfluidity and their related susceptibilities show that, in contrast with
the 2+1 dimensional model, the ground-state at high chemical
potential and low temperature is that of a traditional BCS superfluid.   
This conclusion is supported by the direct observation of a gap in the
dispersion relation for $0.5\le\mu a\le0.85$, which at $\mu a=0.8$ is
found to be roughly $15\%$ the size of the vacuum fermion
mass. We also present results of an initial investigation of the
stability of the BCS phase against thermal fluctuations. Finally, we 
discuss the effect of splitting the Fermi surfaces of the pairing
partners by the introduction of a non-zero isospin chemical potential.
\end{abstract}

\maketitle


\section{Introduction}

The ground state of QCD with $N_f\geq3$ 
flavors of light quark at low temperature $T$ and
asymptotically high density is thought to be 
a ``color-flavor locked'' (CFL) state in which
SU(3)$_c\otimes$SU$(N_f)_L\otimes$SU$(N_f)_R\otimes$U(1)$_B$ symmetry
is spontaneously
broken by diquark condensation in the color anti-triplet channel 
to a diagonal SU(3)$_\Delta$, and 
which is thus simultaneously color superconducting, 
superfluid, and chirally broken~\cite{Alford:1998mk,Schafer:1998ef}.
Since diquark condensation is thought to occur at the Fermi surface via a BCS
mechanism, it is accurately described by a self-consistent gap equation in the
asymptotic regime $\mu\to\infty$, with $\mu$ the quark chemical potential
or Fermi energy, where the QCD coupling $g(\mu)$ is
weak~\cite{Rischke:2003mt}. However, the densities
for which analytic predictions of weakly-coupled QCD can be trusted are
much greater 
than the conditions, with $\mu\sim O(1{\rm G e V})$, which are likely to
be physically realisable in our universe at the centres of compact
stars~\cite{Alford:2000sx}. 

In this regime we face the twin problems that
QCD becomes strongly interacting, and that
the most systematic way of computing its properties
non-perturbatively,
namely numerical simulation of lattice gauge theory, cannot help 
because of the ``sign problem''; the lack of positivity of the QCD Euclidean
path integral measure with $\mu\not=0$ 
makes Monte Carlo sampling methods inoperable.
Another complicating factor is that once the density
becomes low enough to make the strange quark mass $m_s$ no longer negligible
compared to $\mu$, then other channels involving pairing between just $u$ and
$d$ may be preferred, leading to a plethora of different possible ground states,
including even crystalline examples~\cite{Rajagopal:2000wf,Alford:2001dt,
Bowers:2002xr}.
Analytic approaches to these questions 
must either use some approximate 
non-perturbative approach such as the instanton liquid~\cite{Rapp:1998zu}, 
or resort to
phenomenological models of the strong interaction such as the Nambu --
Jona-Lasinio (NJL) model~\cite{Alford:1998zt,Berges:1998rc}.  

Generic NJL models contain fermions, to be thought of either as quarks or
baryons, self-interacting via contact four-Fermi terms~\cite{Nambu:1961tp}. 
In a Euclidean metric, the prototype is written in terms of isopinor
fermion fields $\psi, \bar\psi$, as
\begin{equation}
{\cal L}_{NJL}=\bar\psi(\partial{\!\!\!/\,}+m)\psi-\frac{g^2}{2}\left[
(\bar\psi\psi)^2-(\bar\psi\vec\tau\gamma_5\psi)^2\right],
\label{eq:njl}
\end{equation}
which has SU(2)$_L\otimes$SU(2)$_R$ chiral symmetry.
In 3+1 dimensions the
interaction strength $g^2$ has mass dimension -2, implying that the model is
non-renormalisable and must be defined in terms of
some explicit ultra-violet scale
$\Lambda$ (see e.g. \cite{Hands:1998uf}). Since the model has no
gluons, it fails to reproduce the physics of confinement; however, for
sufficiently large $g^2$
the model does exhibit spontaneous
chiral symmetry breaking to SU(2)$_I$, 
resulting in a physical or ``constituent'' fermion
mass scale $\Sigma\sim g^2\Lambda^3\gg m$. The phase structure of the model is
most conveniently studied in the Hartree approximation, 
corresponding to retaining
only those diagrams which, if the number of fermion degrees of freedom were
multiplied by a factor $N$ and the coupling rescaled as $g^2/N$, 
would remain at leading order in $1/N$. At low $T$  it is found that 
for values of chemical potential $\mu_c\sim\Sigma$ there is a transition,
whose order depends on the details of the cutoff, in which
chiral symmetry is restored and baryon density
$n_B=\langle\bar\psi\gamma_0\psi\rangle$
increases sharply from zero
~\cite{Hatsuda:1985eb,Asakawa:1989bq,Klevansky:1992qe}. 
For $\mu>\mu_c$ the NJL model thus
describes a state resembling relativistic ``quark matter.'' Since there is no
short-range repulsion present, the stability of a phase in which
both $n_B$ and $\Sigma$ are simultaneously non-zero, corresponding to ``nuclear
matter,'' is not {\em a priori} clear. 

The color superconducting solutions discussed in the first paragraphs have been
obtained by solving the self-consistent ``gap equation'' for the smallest energy
$2\Delta$ required to excite a quasiparticle 
pair out of the ground state consisting
of a filled Fermi sea. Solutions with $\Delta>0$, implying the instability of a
sharp Fermi surface with respect to formation of a condensate of
Cooper pairs, form the basis of
the BCS description of superconductivity~\cite{Bardeen:1957mv}. 
Such solutions are also characterised
by a Cooper pair or 
{\em diquark condensate} $\langle q q\rangle\not=0$ whose precise
definition will be given below; here it suffices to identify it with the density
of condensed pairs in the ground state. Since the NJL model is not a gauge
theory the corresponding BCS ground state is not superconducting (analogous to
a Higgs vacuum in particle physics), but rather a superfluid, 
in which the global 
U(1)$_B$ baryon number symmetry of (\ref{eq:njl}) is spontaneously broken by 
$\langle q q\rangle\not=0$, which is thus a true order parameter. 

Since to leading order in $1/N$ $\langle
q q\rangle=\Delta=0$,
it is legitimate to query the 
approximate  methods used to find such solutions. Fortunately in this case
it is possible 
to employ numerical lattice methods, because as reviewed below the
lattice-regularised 
NJL model has no sign problem. Initial studies have focused on
the high density phase of the NJL model in 2+1$d$, in part for the obvious
technical advantage of working in reduced dimensionality, and in part for the
conceptual advantage that NJL$_{2+1}$ has an ultra-violet stable fixed point and
hence an interacting continuum limit, so that in principle we need not worry
about the cutoff dependence of the results. Although the simulations reveal
enhanced diquark pairing in the scalar isoscalar channel~\cite{Hands:1998kk}, 
there is no evidence
for the expected BCS signal $\langle q q\rangle\not=0$, $\Delta\not=0$. Instead,
the condensate vanishes non-analytically with external diquark source
strength $j$, and the fermion dispersion relation reveals a sharp Fermi
surface and massless quasiparticles~\cite{Hands:2000gv,Hands:2001aq}.
The fitted values for the Fermi momentum and Fermi
velocity, however, do not
match those of free field theory, and indeed are difficult to account for in
orthodox Fermi liquid theory~\cite{Hands:2003dh}.
The interpretation is that due to its low dimensionality
the system realises superfluidity in an
unconventional Berezinskii-Kosterlitz-Thouless (BKT) mode first invoked to
describe thin helium films~\cite{Nelson}; 
the massless modes due to
phase fluctuations of the would-be condensate field remain strongly fluctuating,
resulting in unbroken U(1)$_B$ symmetry but critical behaviour for all
$\mu>\mu_c$. Metastability of persistent flow is only revealed by the non-trivial
response of the system to a symmetry breaking field with a twist imposed 
across the spatial boundary.

For this reason in the current paper we present results of simulating
NJL$_{3+1}$, with the goal of exposing superfluid behaviour with the more
conventional signal $\langle q q\rangle\not=0$, $\Delta>0$. Our first results
suggesting that $\langle q q\rangle\not=0$ for $\mu>\mu_c$ have
appeared in~\cite{Hands:2002mr}. The details of the
lattice model, the simulation technique, and the observables monitored to
expose this behaviour are reviewed below in Sec.~\ref{sec:II}.
As already stated, in 3+1$d$ the model is non-renormalisable and must be defined
with a cutoff. The bare parameters $g^2$ and $m$ must be chosen so that the
model's properties match those of low energy QCD as closely as possible; in
Sec.~\ref{ssec:parameter} we use a large-$N$ expansion (here $N$ is identified
with the number of ``colors'' $N_c$) to calculate the vacuum
fermion mass $\Sigma_0$ and the constants $f_\pi$ and $m_\pi$,
and find that a reasonable matching can be made with an inverse lattice spacing
$a^{-1}\simeq700$MeV. In Sec.~\ref{sec:phase} we present results for the
phase structure of the model for low $T$, the 
diquark condensate and associated susceptibilities, and in
Sec.~\ref{sec:disprel} 
dispersion relations in the spin-$\frac12$ sector. 
For the first time there is
evidence for a non-vanishing BCS gap from a lattice simulation, with 
$\Delta/\Sigma_0\simeq0.15$, which translates to a gap of
around 60MeV in physical 
units. This is consistent with estimates from self-consistent 
approaches~\cite{Berges:1998rc,Nardulli:2002ma}. 
In Sec.~\ref{sec:finitev} we discuss finite volume effects and finally
in Secs.~\ref{sec:Tneq0} and \ref{sec:mu_Ineq0} present preliminary results for
simulations both with  $T>0$ and with small but non-zero isospin chemical
potential $\mu_I\not=0$, which has the effect of separating the $u$ and $d$
quark Fermi surfaces and hence discouraging $u d$ condensation.
Sec.~\ref{sec:summary} contains a brief summary.


\section{Lattice Model and Parameter Choice}
\label{sec:II}
\subsection{Formulation and symmetries}
\label{ssec:formulation}

The lattice version of the NJL model we employ was first used in a study 
of chiral symmetry restoration at zero chemical potential
in~\cite{Hands:1998uf}. The action, with lattice
spacing $a$ set to unity, reads
\begin{equation}
S_0=S_{f e r m}+S_{b o s}=\sum_{x y}\bar\chi_x M[\Phi]_{x y}\chi_y
+\bar\zeta_x M^*[\Phi]_{x y}\zeta_y+\frac{2}{g^2}
\sum_{\tilde x}\mbox{tr}\Phi^\dagger_{\tilde x}\Phi_{\tilde x},
\label{eq:Slatt}
\end{equation}
where $\chi,\bar\chi,\zeta$ and $\bar\zeta$ are Grassmann-valued staggered isospinor
fermion fields defined on the sites $x$ of a hypercubic lattice, and
$\Phi\equiv\sigma+i\vec\pi.\vec\tau$ is a $2\times2$ matrix of bosonic auxiliary
fields defined on the dual sites $\tilde x$. The kinetic operator $M$ is given
by
\begin{eqnarray}
M_{x y}^{p q}&=&{\frac12}\delta^{p q}\left[
e^\mu\delta_{y x+\hat0}-e^{-\mu}\delta_{y x-\hat0}+\sum_{\nu=1,2,3}\eta_\nu(x)
(\delta_{y x+\hat\nu}-\delta_{y x-\hat\nu})+2m\delta_{x y}\right]\nonumber\\
&+&{\frac{1}{16}}\delta_{x y}\sum_{\langle\tilde x,x\rangle}\lt(\sigma(\tilde
x)\delta^{p q}+i\varepsilon(x)\vec\pi(\tilde x).\vec\tau^{p q}\rt)
\label{eq:M}
\end{eqnarray}
such that the parameters are bare fermion mass $m$, coupling $g^2$ (with mass
dimension -2) and baryon chemical potential
$\mu$. The symbol $\langle\tilde x,x\rangle$ denotes the set of 16 dual sites
adjacent to $x$. 
The Pauli matrices acting on isospin indices $p,q$ are normalised so that
$\mbox{tr}\tau_i\tau_j=2\delta_{i j}$. The phase factors
$\eta_\nu(x)=(-1)^{x_0+\cdots+x_{\nu-1}}$ and
$\varepsilon(x)=(-1)^{x_0+x_1+x_2+x_3}$ ensure that fermions with the correct
Lorentz covariance properties emerge in the continuum limit, and that in the
limit $m\to0$ the action (\ref{eq:Slatt}) has a global
SU(2)$_L\otimes$SU(2)$_R$ invariance under
\begin{eqnarray}
\chi\mapsto({\cal P}_e U+{\cal P}_o V)\chi\;\;&;&\;\;
\bar\chi\mapsto\bar\chi({\cal P}_e V^\dagger+{\cal P}_o U^\dagger)\nonumber\\
\zeta\mapsto({\cal P}_e U^*+{\cal P}_o V^*)\zeta\;\;&;&\;\;
\bar\zeta\mapsto\bar\zeta({\cal P}_e V^{\dagger*}+{\cal
P}_o U^{\dagger*})\nonumber\\
\Phi&\mapsto&V\Phi U^\dagger\;\;\;\;\mbox{with}\;\;\;U,V\in\mbox{SU(2)}.
\label{eq:su2xsu2}
\end{eqnarray}
The even/odd projectors are given by ${\cal
P}_{e/o}(x)={\frac12}(1\pm\varepsilon(x))$. In addition the action has a U(1)$_B$
invariance under baryon number rotations:
\begin{equation}
\chi,\zeta\mapsto
e^{i\alpha}\chi,\zeta\;\;\;\;\;\bar\chi,\bar\zeta\mapsto\bar\chi,\bar\zeta
e^{-i\alpha}.
\label{eq:u1b}
\end{equation}

The auxiliary fields $\sigma$ and $\vec\pi$ 
can be integrated out to yield an action
written solely in terms of fermion fields $\chi$ and $\zeta$.
In the continuum limit $a\partial\to0$ lattice fermion
doubling leads to a physical content for 
the model of eight fermion species (four described by $\chi$, four by $\zeta$)
with self-interactions resembling those of the continuum NJL model (but with 
terms of the form $\bar\chi\chi\bar\zeta\zeta$ differing in detail
from those of the form $\bar\chi\chi\bar\chi\chi$),  
resulting in an additional approximate U(4)$\otimes$U(4) symmetry 
which is violated by
terms of $O(a)$. Further details are given in~\cite{Hands:1998uf}. In what
follows we will somewhat arbitrarily refer to these degrees of freedom as {\em
colors} and hence define $N_c=8$; this is in distinction to the explicit 
flavor or isospin symmetry (\ref{eq:su2xsu2}) which gives rise to $N_f=2$.

In order to examine issues associated with diquark pairing, it is necessary to
introduce a source term into the action. A suitable term, invariant 
under (\ref{eq:su2xsu2}) but violating baryon number conservation (\ref{eq:u1b}),
is:
\begin{equation}
S_{source}=\sum_x {\frac j2}(\chi^{tr}_x\tau_2\chi_x+\zeta^{tr}_x\tau_2\zeta_x)
+{\frac{\bar\jmath}2}(\bar\chi_x\tau_2\bar\chi^{tr}_x
+\bar\zeta_x\tau_2\bar\zeta^{tr}_x),
\end{equation}
where in this paper we will consider $j,\bar\jmath$ real and positive.
With this addition the partition function follows from integrating over the 
Grassmann degrees of freedom:
\begin{eqnarray}
Z[j,\bar\jmath]&=&\int D\chi D\bar\chi D\zeta D\bar\zeta D\Phi
\exp-(S_{f e r m}+S_{b o s}+S_{source})\nonumber\\
&=& \int D\Phi\;\mbox{Pf}^{N_c/4}({\cal A}[\Phi,j,\bar\jmath])
e^{-S_{b o s}[\Phi]}
\label{eq:Z}
\end{eqnarray}
with the antisymmetric matrix ${\cal A}(j,\bar\jmath)$, which acts on bispinors
$(\bar\chi, \chi^{tr})$ (and ${\cal A}^*(-j,-\bar\jmath)$ 
on $(\bar\zeta,\zeta^{tr})$), given by
\begin{equation}
{\cal A}=\left(
\begin{matrix}\bar\jmath\tau_2&M\cr-M^{tr}&j\tau_2\cr
\end{matrix}\right).
\end{equation}
The reality of $\mbox{Pf}{\cal A}=\sqrt{\mbox{det}{\cal A}}$ follows
by noting the identity
$\tau_2M\tau_2=M^*$~\cite{Hands:2001aq}. In fact, 
$\mbox{Pf}{\cal A}$ can also be shown to be positive 
definite~\cite{Hands:2001aq,Hands:2000ei}, implying that $Z$ can
also be defined for a single power of the pfaffian, (i.e. with the fields
$\zeta,\bar\zeta$ omitted), corresponding to $N_c=4$.
However, the smallest value of $N_c$ permitted by the exact hybrid Monte Carlo 
algorithm (see Sec.~\ref{ssec:simulation}) is 8.

\subsection{Observables}
\label{ssec:observables}
 
Here we list the various observables monitored in the simulations. For
simplicity we restrict our attention to those constructed from the 
$\chi,\bar\chi$ fields; their equivalents in the $\zeta$ sector are easily
found. Angled brackets denote averages taken with respect to the measure defined
by (\ref{eq:Z}).

The chiral condensate $\langle\bar\chi\chi\rangle$ is given by
\begin{equation}
\langle\bar\chi\chi\rangle={\frac1V}{\frac{\partial\ln Z}{\partial m}}=
{\frac1{2V}}\left\langle\mbox{tr}\biggl(
\begin{matrix}&\openone\cr-\openone&\cr\end{matrix}\biggr)
{\cal A}^{-1}\right\rangle.
\label{eq:chibarchi}
\end{equation}
To leading order in a large-$N_c$ expansion the chiral condensate is
proportional to the expectation value of the scalar field
$\Sigma\equiv\langle\sigma\rangle=\frac{g^2}{2}
\langle\bar\chi\chi\rangle$, which in the
chiral limit also defines the physical or {\em constituent} fermion mass.
The baryon charge density $n_B$ per flavor is given by 
\begin{equation}
n_B=\frac{1}{2V}{\frac{\partial\ln Z}{\partial\mu}}=
{\frac1{8V}}\left\langle\mbox{tr}\biggl(
\begin{matrix}&e^\mu\delta_{y x+\hat0}+e^{-\mu}\delta_{y x-\hat0}\cr
-e^\mu\delta_{y x-\hat0}-e^{-\mu}\delta_{y x+\hat0}&\cr\end{matrix}\biggr)
{\cal A}^{-1}\right\rangle.
\label{eq:nB}
\end{equation}
In the diquark sector it is convenient to define operators
\begin{equation}
q q_{\pm}(x)={\frac14}[\chi^{tr}_x\tau_2\chi_x\pm\bar\chi_x\tau_2\bar\chi^{tr}_x]
\end{equation}
and corresponding source strengths $j_\pm=j\pm\bar\jmath$.
The condensates are then
\begin{equation}
\langle q q_\pm\rangle={\frac1V}{\frac{\partial\ln Z}{\partial j_\pm}}=
{\frac1{4V}}\left\langle\mbox{tr}
\biggl(\begin{matrix}\pm\tau_2&\cr&\tau_2\end{matrix}\biggr){\cal A}^{-1}
\right\rangle.
\label{eq:qq}
\end{equation}
In the continuum limit, the corresponding operators written in terms of spinors
$\psi,\bar\psi$ 
with 4 spin, 2 flavor and 4 color components are~\cite{Hands:2000ei}
\begin{equation}
q q_\pm\propto\psi^{tr}C\gamma_5\otimes\tau_2\otimes(C\gamma_5)^*\psi\pm
\bar\psi C\gamma_5\otimes\tau_2\otimes(C\gamma_5)^*\bar\psi^{tr},
\end{equation}
with $C$ the charge conjugation matrix satisfying $C\gamma_\mu
C^{-1}=-\gamma_\mu^*$. The condensate wavefunction is thus scalar isoscalar, but
with a non-trivial (and for us uninteresting) 
variation under ``color'' rotations, and
manifestly antisymmetric as required by the exclusion principle.
We will also consider the diquark susceptibilities
\begin{equation}
\chi_\pm\equiv{\frac{\partial\langle q q_\pm\rangle}{\partial j_\pm}}=
\sum_x\langle q q_\pm(0)q q_\pm(x)\rangle
\label{eq:susc}
\end{equation}
which are constrained by identities analogous to the axial Ward identity
for the pion propagator; a particularly useful form in the $j_-\to0$ limit reads
\begin{equation}
\chi_-\vert_{j_-=0}={\frac{\langle q q_+\rangle}{j_+}}.
\label{eq:Ward}
\end{equation}
When U(1)$_B$ is spontaneously broken by the formation of a condensate
$\langle q q_+\rangle\not=0$, therefore, $q q_-$ interpolates the
resulting Goldstone mode, 
whose masslessness as $j_+\to0$ is guaranteed by
(\ref{eq:Ward}). Physically the Goldstone mode is responsible for long-range
interactions between vortices in the superfluid phase, and at non-zero
$T$ for propagating
wave excitations of local energy density known as second sound.

In order to study the spectrum in the spin$-{\frac12}$ sector we need the 
{\em Gor'kov} propagator ${\cal G}={\cal A}^{-1}$. It can be written as
\begin{equation}
{\cal G}_{x y}=\left(\begin{matrix} A_{x y}&N_{x y}\cr\bar N_{x y}&\bar A_{x y}\cr
\end{matrix}\right)
\label{eq:gorkov}
\end{equation}
where each entry is a $2\times2$ matrix in isospace. The notation indicates
both {\em normal} $\langle q(x)\bar q(y)\rangle$ and {\em anomalous} $\langle
q(x)q(y)\rangle$ components: on a finite system $A$ and $\bar A$ vanish in the 
limit $j,\bar\jmath\to0$, and we recover the usual fermion and anti-fermion
propagators as $N$ and $\bar N$ respectively. The number of independent
components is constrained by the identities $N_{22}\equiv N_{11}^*$,
$N_{21}\equiv-N_{12}^*$, $A_{22}\equiv-A_{11}^*$ and $A_{21}\equiv 
A_{12}^*$, so that
${\cal G}$ can be reconstructed using just two conjugate gradient inversions of
${\cal A}$. In addition, isospin 
and time-reversal symmetries dictate that after averaging over
configurations the only independent non-vanishing components are
$\mbox{Re}\overline N_{11}$ and ${\rm I m}
A_{12}$, which henceforth we will refer to simply as $N$ and $A$ respectively.

\subsection{Phenomenological Parameter Choice}
\label{ssec:parameter}

As previously discussed, in a number of dimensions $d\ge4$, the
continuum NJL model is non-renormalisable, which implies that
the lattice model becomes trivial in the continuum
limit~\cite{Klevansky:1992qe, Hands:1998uf}.  This means one
must choose a fixed lattice spacing $a$, corresponding to a fixed inverse
coupling constant $\beta=\frac{a^2}{g^2}$, by fitting the
model's parameters to low energy, vacuum phenomenology.
Employing methods outlined
in~\cite{Klevansky:1992qe}, we calculate the ratio between the pion decay rate
$f_\pi$ and the constituent quark mass, i.e. the dynamical 
mass of the
quark in the chirally broken phase.  By fitting to phenomenological
values we may extract $\beta$ as a function of the model's only
other free parameter, the current quark mass $m$. Finally, calculating
and fitting the pion mass 
$m_\pi$ allows us to fix $m$, and hence $\beta$.  We take advantage
of the fact that a perturbative expansion in $1/N_c$ is possible in
four-Fermi theories by calculating quantities analytically to
leading order 
in $1/N_c$, the Hartree approximation.  Feynman diagrams are
evaluated using staggered quark propagators defined on an $L_s^3\times
L_t/2^4$ Euclidean blocked lattice
with periodic boundary conditions in spatial dimensions and
anti-periodic boundary conditions in the temporal dimension. The
integrals over loop momenta are evaluated as sums over momentum modes.

Let us first calculate the gap equation, the fermion self-interaction,
to leading order in $1/N_c$. For sufficiently strong coupling
$g^2>g_c^2$ the scalar auxiliary field $\sigma$ develops a spontaneous
vacuum expectation value $\Sigma$, which in the chiral limit can be
identified with the constituent fermion mass  and is given by
\be\begin{array}{c c c c c}
\Sigma&=&
\begin{picture}(30,20)(0,0)
\Line(0,-20)(0,20)
\ArrowArc(10,0)(10,181,180)
\Vertex(0,0){2}
\end{picture}
&=&{\displaystyle \frac{N_c N_f \Sigma^*}{\beta}
\int^\frac{\pi}{2}_{-\frac{\pi}{2}}\frac{{\rm d}^4p}{\lt(2\pi\rt)^4}
\frac{1}{\tilde p^2 +\lt(\Sigma^*a\rt)^2}},
\end{array}
\label{eq:gapeqn}\ee
where $\tilde p^2=\sum_{\nu=0}^3\sin^2 p_\nu$ is the dimensionless squared
loop momentum,  $N_c$ and $N_f$ are the number of flavors and colors
respectively, and $\beta=a^2/g^2$ is the dimensionless inverse
coupling constant. In this section we shall distinguish the
expectation value of the scalar field $\Sigma$ from the
constituent mass, which to leading order in $(1/N_c)$ we define as
$\Sigma^*\equiv\Sigma+m$; elsewhere in this paper the two shall be
used interchangeably and both denoted by $\Sigma$.

$f_\pi$ is calculated from the vacuum to one-pion axial-vector matrix element
\be
\begin{picture}(300,30)(-140,30)
\ArrowArc(43,5)(50,30,150)
\ArrowArc(43,55)(50,210,330)
\Vertex(0,30){2}
\DashArrowLine(135,30)(85,30){1}
\Vertex(85,30){3}
\Text(110,35)[b]{$k a^{-1}$}
\Text(43,57.5)[b]{$(p+k/2)a^{-1}$}
\Text(43,2.5)[t]{$(p-k/2)a^{-1}$}
\Text(-5,30)[r]{$\lt<0\lt|J_{i5\mu}(k)\rt|\pi_j\rt> =
i\gamma_\mu\gamma_5\frac{\tau_j}{2}$}
\Text(80,30)[r]{$i g_{\pi q q}\gamma_5\tau_i$}
\SetColor{White}
\Vertex(85,30){2}
\end{picture}.
\vspace{1cm}
\ee
Translating this diagram, noting that the pion-quark-quark coupling 
$g_{\pi q q}\sim g/\surd N_c$, and
taking the $k\to0$ limit we find 
\be
\frac{f_\pi}{\Sigma^*}=\frac{\displaystyle \sqrt{2N_c N_f}
\int^{\frac{\pi}{2}}_{-\frac{\pi}{2}}{\frac{{\rm d}^4p}{\lt(2\pi\rt)^4}
\frac{\cos{2p_\nu}}{\lt[\tilde p^2+\lt(\Sigma^*a\rt)^2
\rt]^2}}}{\displaystyle 
\lt(\int^{\frac{\pi}{2}}_{-\frac{\pi}{2}}{\frac{{\rm d}^4p}{\lt(2\pi\rt)^4}
\frac{1}{\lt[\tilde p^2+\lt(\Sigma^* a\rt)^2\rt]^2}}\rt)
^{\frac{1}{2}}}.
\label{eq:fpifinal}
\ee

In order to check that this form is sensible we choose to examine the
continuum limit by extracting the leading order behaviour of
(\ref{eq:fpifinal})  as the dimensionless quark mass $\Sigma^* a$ is
reduced to zero. This is done by the introduction of a dimensionless 
hyper-spherical cut-off $\delta$ which splits the loop momenta into
two regions, one with $\lt|p\rt|>\delta$ and one with
$\lt|p\rt|\leq\delta$.  As $\delta\to0$, the inner region picks up the
continuum behaviour, whilst the outer region yields the
finite terms relevant to lattice perturbation
theory~\cite{Karsten:1981wd}.  Ignoring these terms and taking
$\Sigma^* a\to0$
we pick up a leading contribution 
$f_\pi/\Sigma^*\sim\sqrt{N_c N_f\ln(\delta/\Sigma^* a)/4\pi^2}$.
Although this quantity is logarithmically divergent, we know that the
integral between $\pi/2$ and $-\pi/2$ is both finite and independent of
$\delta$, and that the transition between the two regions of
integration is smooth. This means that the $\ln\delta$ term must be
cancelled out by a similar term in the outer region and the 
continuum behaviour of $f_\pi/\Sigma^*$ is thus given by
\be
\frac{f_\pi}{\Sigma^*} \sim \sqrt{\frac{N_c N_f}{\lt(2\pi\rt)^2}
\ln{\frac{1}{\Sigma^*a}}}.
\ee
This is the same behaviour found as $\Sigma^*/\Lambda\to0$ for the
regularisation schemes employed in~\cite{Klevansky:1992qe}, namely
$3d$-momentum, $4d$-momentum, and real-time cut-offs as well as the
Pauli-Villars scheme. 

By calculating (\ref{eq:fpifinal}) in the infinite volume limit, 
fitting $f_\pi$ to its experimental value of 93MeV and $\Sigma^*$ to a
reasonable 400MeV we are able to extract the dimensionless quark mass
$\Sigma^*a=0.557$, such that to leading order in $1/N_c$ the lattice spacing
$a=(720{\rm MeV})^{-1}\sim0.3{\rm fm}$. Solving the gap equation
(\ref{eq:gapeqn}) with this value  for the mass we find that
$\beta\Sigma a=0.273$. Using the identity $\Sigma^*\equiv
m+\Sigma$ we deduce a relationship between the bare
quark mass and the inverse coupling:
\be
\beta=\frac{0.273}{(0.557-m a)}.
\label{eq:beta(m0)}
\ee  
Finally, in order to fix the bare quark mass and hence the coupling,
we need to fit one more phenomenological
observable. Again following~\cite{Klevansky:1992qe}, we calculate the mass of
the pion $m_\pi$ by solving the self-consistent equation
\be
4\frac{m}{\Sigma^*}=2N_c N_f\frac{(m_\pi a)^2}{\beta}I,
\ee
where $I$ represents the integral
\be
I=
\int^{\frac{\pi}{2}}_{-\frac{\pi}{2}}
{\frac{{\rm d}^4p}{\lt(2\pi\rt)^4}
\frac{1} {\lt[\tilde p^2_++({\Sigma^*}a)^2 \rt]
\lt[\tilde p^2_-+({\Sigma^*}a)^2
\rt]}},
\ee
and $\tilde p^2_\pm=\sum_{\nu=0}^3\sin^2(p\pm i m_\pi a)_\nu$.
Setting $m_\pi$ to a phenomenologically reasonable 138MeV and
demanding that (\ref{eq:beta(m0)}) is satisfied fixes the bare mass to
$ma=0.006$ and the inverse coupling to $\beta=0.495$.  Table
\ref{tab:regsum} contains a summary of the fits made and
parameters extracted. 

\begin{table}[ht]
\centering
\begin{tabular}{|l|l|}
\hline\hline
Phenomenological & Lattice Parameters\\ Observables Fitted &
Extracted\\
\hline
$\Sigma^*=400$MeV & $ma=0.006$\\ $f_\pi=93$MeV & $\beta=0.495$\\
$m_\pi=138$MeV & $a^{-1}=720$MeV\\
\hline\hline
\end{tabular}
\caption{Summary of large-$N_c$ parameter fits.}
\label{tab:regsum}
\end{table}

\subsection{The Simulation}
\label{ssec:simulation}

We chose to perform 
the numerical simulation of the path integral defined by (\ref{eq:Z}) in the
{\em partially quenched} limit $j=\bar\jmath=0$, for two reasons. The first is
technical; in this limit $\mbox{Pf}^2({\cal A})=\mbox{det}MM^\dagger$, so that
there is no obstruction to using the hybrid Monte Carlo (HMC)
algorithm~\cite{Duane:1987de}, which is ``exact'' in the sense that
there is no 
systematic 
error due to non-zero time step, a reassuring feature whenever a previously
unexplored phase is to
be studied. The second is pragmatic; it enables many values of $j$ to
be studied 
with a single run at fixed $g^2$ and $\mu$, thus saving much computer time. Our
previous studies in 2+1$d$~\cite{Hands:2000gv,Hands:2001aq} used both partially
quenched and full pfaffian simulations, and found essentially no difference in
the results.

The simulations were run on a variety of lattice sizes and values of $\mu$.
In a typical simulation, we generated 500 equilibrated
configurations separated by HMC trajectories of mean length 1.0. The
time step required to achieve an acceptance rate $\sim80\%$  was found
to decrease with increasing lattice size. For larger lattices, where
the optimal time step was smaller than $1/25$, we also tuned the number of
colors  $N_c'\agt N_c=8$ used during the generation of molecular
dynamic trajectories; in tuning the number of colors, which is 
``renormalised'' by the discretisation of the trajectory, 
one can increase the
acceptance without increasing the number of integration steps used. 
In the exact accept/reject step, and hence the integral over all
configurations, $N_c\equiv8$.  
Measurements were carried out on every second configuration, in all
cases with $j=\bar\jmath\Rightarrow j_-=0$. In the rest of the 
paper we will quote all results in terms of $j_+$, henceforth referred
to as $j$. 

The reader may be wondering why HMC simulations of the NJL model with
$\mu\not=0$ 
are possible, whereas they are well-known to be ineffective for QCD. In
both cases the algorithm reproduces 
as path integral measure the positive definite
$\mbox{det}MM^\dagger$. For QCD $M$ describes a color triplet of
quarks $q$, $M^\dagger$ a color anti-triplet of conjugate quarks $q^c$, thus
permitting gauge singlet baryonic $q q^c$ bound states. The lightest such state
can be shown to be degenerate with the Goldstone 
pion associated with chiral symmetry breaking, which means that baryonic matter
is induced into the ground state at an onset $\mu_{o HMC}\sim O(m_\pi/2)$, in
contradiction to the physical expectation $\mu_{o QCD}\sim
O(m_{nucleon}/3)$~\cite{Gocksch:1988ha,Stephanov:1996ki}. 
In NJL-like models, by contrast, in the large-$N_c$ limit 
the Goldstone pole is saturated by disconnected
$q\bar q$ bubbles, which cannot contribute in the $q q^c$
channels~\cite{Barbour:1999mc}; hence the HMC onset occurs at the
constituent scale 
$\mu_o\sim\Sigma$ as expected, resulting in a ground state of degenerate 
fermion degrees of freedom and hence a Fermi surface. For the continuum
model studied here
$q$ and
$q^c$ have identical quantum numbers and are hence indistinguishable; however
the HMC approach also yields physically reasonable behaviour even in models
where the chiral symmetry group is U(1) rather than SU(2)$\otimes$SU(2) and this
is not the case~\cite{Barbour:1999mc}.

The observables defined as matrix traces in
equations~(\ref{eq:chibarchi},\ref{eq:nB},\ref{eq:qq}) 
may be estimated on a particular 
$\Phi$ configuration by calculating the bilinear $\eta^\dagger\Gamma{\cal
A}^{-1}\eta$ using a conjugate gradient algorithm with a stochastic source vector
$\eta$ satisfying $\overline{\eta^\dagger\eta}\propto\delta_{x y}\delta^{p q}$;
during each measurement we used a set of 5 such vectors.
The Gor'kov propagator ${\cal G}$ is similarly calculated but using this time a
point source. Special care is needed for susceptibilities, which contain
contributions from both
connected and disconnected fermion line diagrams, and hence must be calculated
using both methods via expressions such as
\begin{equation}
\chi=\langle
\overline{\eta^{i\dagger}\Gamma{\cal A}^{-1}\eta^i\eta^{j\dagger}\Gamma
{\cal A}^{-1}\eta^j}\rangle-\langle\overline{\eta^{i\dagger}\Gamma{\cal A}^{-1}\eta^i}
\rangle^2+
\sum_x\langle\mbox{tr}{\cal G}_{0x}\Gamma{\cal
G}_{0x}^{tr}\Gamma\rangle.
\end{equation}
We label these two contributions the disconnected and connected parts
of $\chi$ respectively. 


\section{Zero Temperature Phase Structure}
\label{sec:phase}

\subsection{Chiral Symmetry Restoration}
\label{ssec:eofs} 

In this section we investigate the nature of the chiral restoration
transition with 
our phenomenologically motivated parameter set, since the order of this
transition is found to be sensitive to the choice of parameters
employed~\cite{Klevansky:1992qe}. We measure the chiral
condensate $\lt<\chibar\chi\rt>$ and baryon number density $n_B$
on $V=L_s^3\times L_t$ lattices with $L_s=L_t=12$, 16 and
20, and for various chemical potentials $\mu a$ between 0.0 and 1.2.
Henceforth all dimensionful quantities will be quoted in lattice units $a=1$.
The quantum average and statistical errors of the measured quantities
are calculated using a jackknife estimate and the results extrapolated
linearly to the infinite volume limit $V^{-1}\to0$.

\begin{figure}[h]
\centering
\includegraphics[width=12cm]{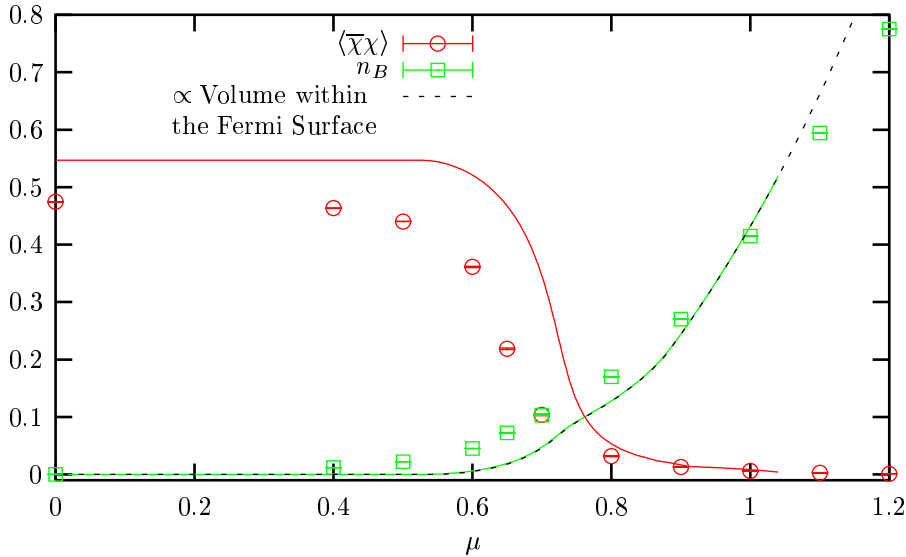}
\caption{Chiral condensate and number density extrapolated to
$V^{-1}\to0$ as functions of $\mu$,
 showing both the large-$N_c$ solution
(solid curve) and lattice results (points). The dashed curve
is proportional to the volume of the Brillouin zone bounded by the
large-$N_c$ Fermi surface.}
\label{fig:EofS}
\end{figure}

Our results are presented in Fig.~\ref{fig:EofS}.
In order to compare the lattice data (points) with perturbative results, 
both $\left<\overline\chi\chi\right>$ and $n_B$ are calculated to leading order
in $1/N_c$ (solid curves), corresponding to a mean-field theory
in which the scalar field $\sigma=\Sigma$ on every dual lattice site 
and the auxiliary pseudoscalars $\pi_i$ are exactly zero; in this
limit the condensate is given by $\left<\overline\chi\chi\right>=
\frac{2}{g^2}\Sigma\equiv\frac{2}{g^2}\left<\sigma\right>$.
At $\mu=0$, the large-$N_c$ solution predicts a non-zero condensate and
zero baryon density, corresponding to a vacuum with broken chiral
symmetry. As $\mu$ is increased the system passes
into a phase in which chiral symmetry is approximately
restored and matter begins to build up, with a pronounced ``kink''
at $\mu_c\simeq0.75$. In particular, $n_B$ is 
seen to be closely related to the volume of momentum space enclosed by the
Fermi surface defined by the Fermi momentum $\vec k_F$ given for free
fermions by 
\be
\sinh^2\mu=\sum_{i=1}^3\sin^2{k_{F i}}+\Sigma^2 
\label{eq:Surface}
\ee
and plotted in the large-$N_c$ limit as the dashed line.
This implies that in the continuum $n_B\propto k_F^3$ as one would
expect, whilst on a lattice $n_B$ saturates at the value 1.0 as 
$\mu\to\infty$.

The lattice data agree qualitatively with our analytic solution, 
although for these data both $\left<\overline\chi\chi\right>$ and
$\mu_o$ are roughly 
15\% smaller, an effect we attribute to ${O}(1/N_c)$ corrections.
The transition at $\mu\sim\mu_c$ has the appearance of a crossover, and may thus
be compatible with a second order transition in the chiral limit $m\to0$. If
this is the case the region with $n_B>0$ for $\mu<\mu_c$ could be associated with a
``nuclear matter'' phase, since 
if, as $T\to0$,  $\Sigma$ falls below its vacuum value $\Sigma_0$
for $\mu<\mu_c$, the only possible physical agency is $n_B>0$ 
(a mixed phase of quark matter droplets in vacuum at constant pressure
is not consistent with
thermodynamic equilibrium unless there are at least two conserved quantum
numbers present \cite{Glendenning:1992vb}).
This behaviour is in marked contrast with that of 
the 2+1$d$ model in which the transition is strongly
first-order and baryonic matter has chiral symmetry 
restored at any density~\cite{Hands:1998kk,Hands:2001aq,Kogut:1999um}.
Finally, it is illustrative to convert these densities into physical units.
The large-$N_c$ estimate $a^{-1}=720$MeV translates the lattice point
$(\mu a,n_Ba^3)\simeq(0.65,0.072)$ into $\mu=470$MeV, $n_B=3.5\mbox{fm}^{-3}$.
Bearing in mind that due to species doubling, in a cube of volume $(2a)^3$ $\chi$ 
describes two spin and four color components of a continuum fermion, 
we deduce
a total physical density of $0.88N_c N_f\mbox{fm}^{-3}$, to be
compared with the  nuclear matter onset $\mu_q\simeq320$MeV,
$n_q\simeq0.15N_c\mbox{fm}^{-3}$. 

\subsection{Diquark Condensation}
\label{ssec:qq}

The main purpose of this study is to
determine the nature of the high density phase in which chiral
symmetry is approximately restored. In
particular, in order to explore the possibility of a U(1)$_B$-violating
BCS phase, we study the diquark order parameters (\ref{eq:qq}) and
their susceptibilities (\ref{eq:susc}) as functions of
chemical potential. On their own, the susceptibilities are of limited
importance. In conjunction with the Ward identity (\ref{eq:Ward})
however, their ratio
\begin{equation}
R=-\frac{\chi_+}{\chi_-}
\label{eq:R}
\end{equation}
provides an important tool to distinguish between phases in
which U(1)$_B$ 
symmetry is either manifest or broken. With manifest symmetry, and
in the limit $j\to0$, these two susceptibilities should be identical
up to a sign factor, and the ratio should equal 1. If the symmetry is
broken, however, the Ward identity predicts that the Goldstone mode
$\chi_-$ should diverge as $j\to0$ and hence $R$ should vanish. 

As stated in Sec.~\ref{ssec:simulation}, the disconnected terms of
(\ref{eq:susc}) are calculated via the use of multiple noise vector
estimation, whilst the symmetry constraints discussed after
(\ref{eq:gorkov}) imply that the connected terms are given by 
\be
\chi^{con}_\pm=\frac{1}{4}\lt(-\lt[\lt|\Go_{11}\rt|^2+\lt|\Go_{21}\rt|^2
+\lt|\Go_{33}\rt|^2+\lt|\Go_{43}\rt|^2\rt]
\pm\lt[\lt|\Go_{13}\rt|^2+\lt|\Go_{23}\rt|^2
+\lt|\Go_{31}\rt|^2+\lt|\Go_{41}\rt|^2\rt]\rt),
\ee 
evaluated between a random point source and the point $x$.

The susceptibilities are measured and $R$ calculated on the
aforementioned lattice sizes and for various values of $\mu$, with the
diquark source $j$ varying from 0.1 to 1.0 during each set of
measurements. It is interesting to note that
although in most cases the disconnected contributions are found to be
consistent with zero, in the low $\mu$ phase with large $j$,
$\chi_+^{dis}$ can be up to 10-20\% the magnitude of
$\chi_+^{con}$. In contrast with the NJL model in
2+1$d$~\cite{Hands:2001aq}, therefore, we cannot assume that
$\chi_+\simeq\chi_+^{con}$. 

\begin{figure}[ht]
\centering
\includegraphics [width=5.5cm]{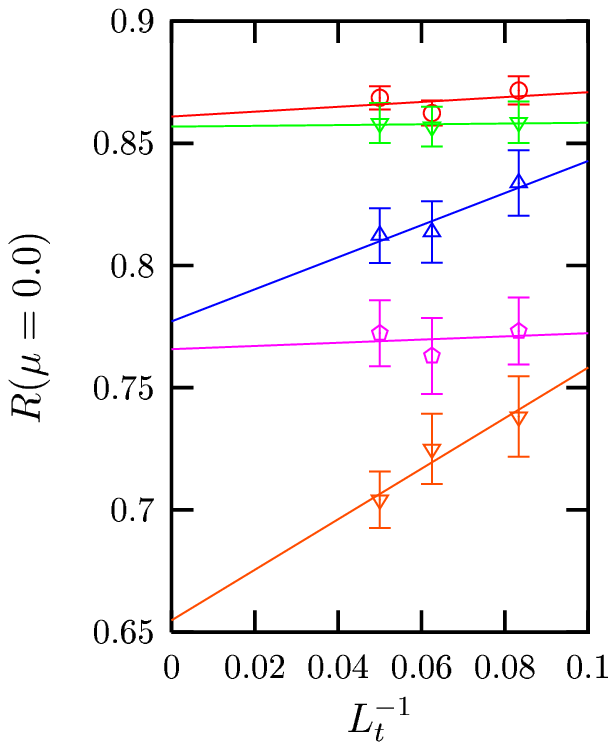}
\includegraphics [width=5.5cm]{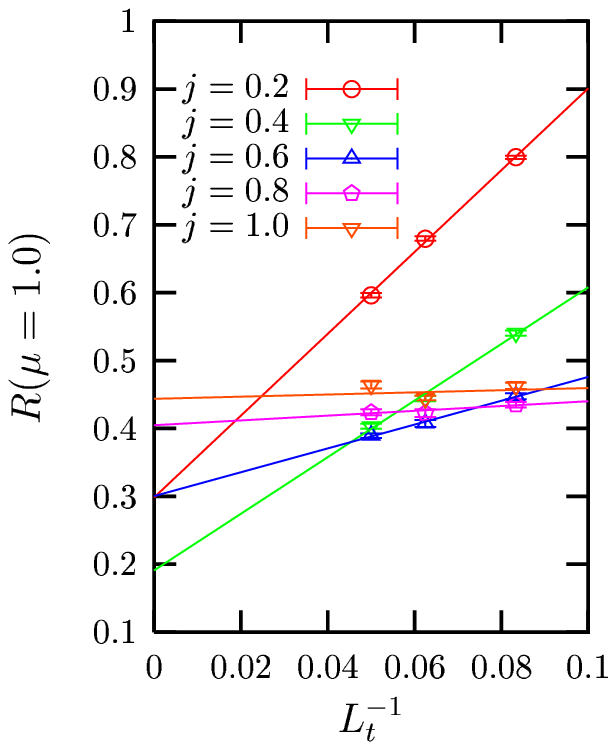}
\caption{Susceptibility ratio $R$ at $\mu=0.0$ and 1.0 on various
lattice sizes.}
\label{fig:Ltextrap}
\end{figure}

An interesting empirical observation is that whilst the observables
measured in Sec.~\ref{ssec:eofs} were found to scale linearly with
the inverse volume of the lattice, observables in the diquark sector
appear to scale linearly with the inverse temporal extent,
corresponding to the temperature of the system.
Accordingly, the ratio $R$ is extrapolated linearly to the limit
$L_t^{-1}\to0$, which as in~\cite{Hands:2001aq} is found to give a reasonable 
description of the data.
An example of the quality of the fits is illustrated in
Fig.~\ref{fig:Ltextrap}. 

\begin{figure}[h]
\centering
\includegraphics[width=12cm]{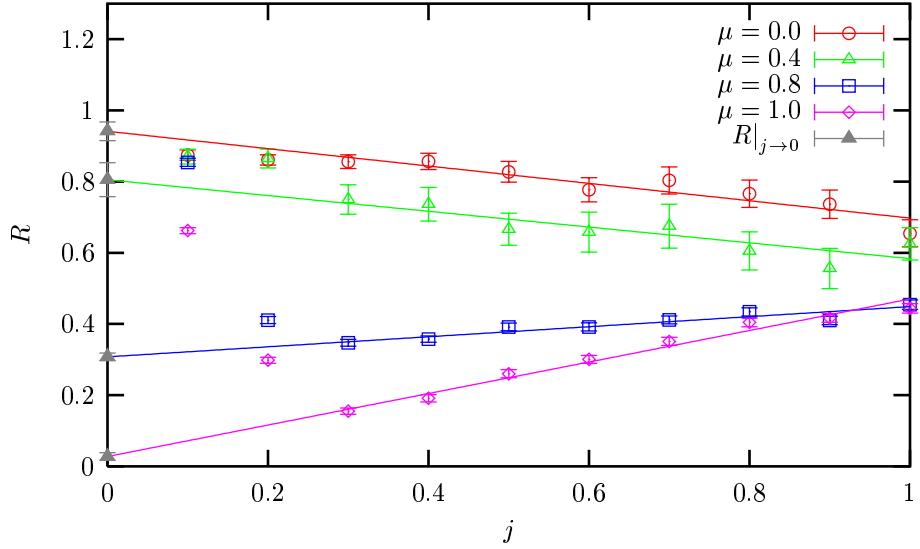}
\caption{$R$ extrapolated to $j\to0$ for various values of $\mu$.}
\label{fig:Rvsj}
\end{figure}
Figure~\ref{fig:Rvsj} shows this extrapolated data plotted as a function of
$j$ for various values of $\mu$, as well as the results of a linear
extrapolation to $j\to0$. One immediately notices that whilst this
extrapolation appears plausible for $j\ge0.3$, the data
at lower values of $j$ in the high $\mu$ phase diverge rapidly from
the linear trend. The fact that this effect increases systematically with
increasing $\mu$ and decreasing $j$ suggests that its origin is some
systematic effect not considered thusfar; indeed, the study
presented in Sec.~\ref{sec:Tneq0} shows this to be due to residual
finite temperature effects.
For this reason, we believe that we are justified in disregarding
data with $j\le0.2$ when taking the $j\to0$ limit. 
With this omission Fig.~\ref{fig:Rvsj} shows that for $\mu=0$ a linear
fit is consistent 
with a ratio of $R\approx 1$, corresponding to a manifest baryon
number symmetry as one would expect in the vacuum. At $\mu=1.0$, however,
$R\approx0$ suggesting that U(1)$_B$ symmetry is broken.

\begin{figure}[h]
\centering
\includegraphics[width=12cm]{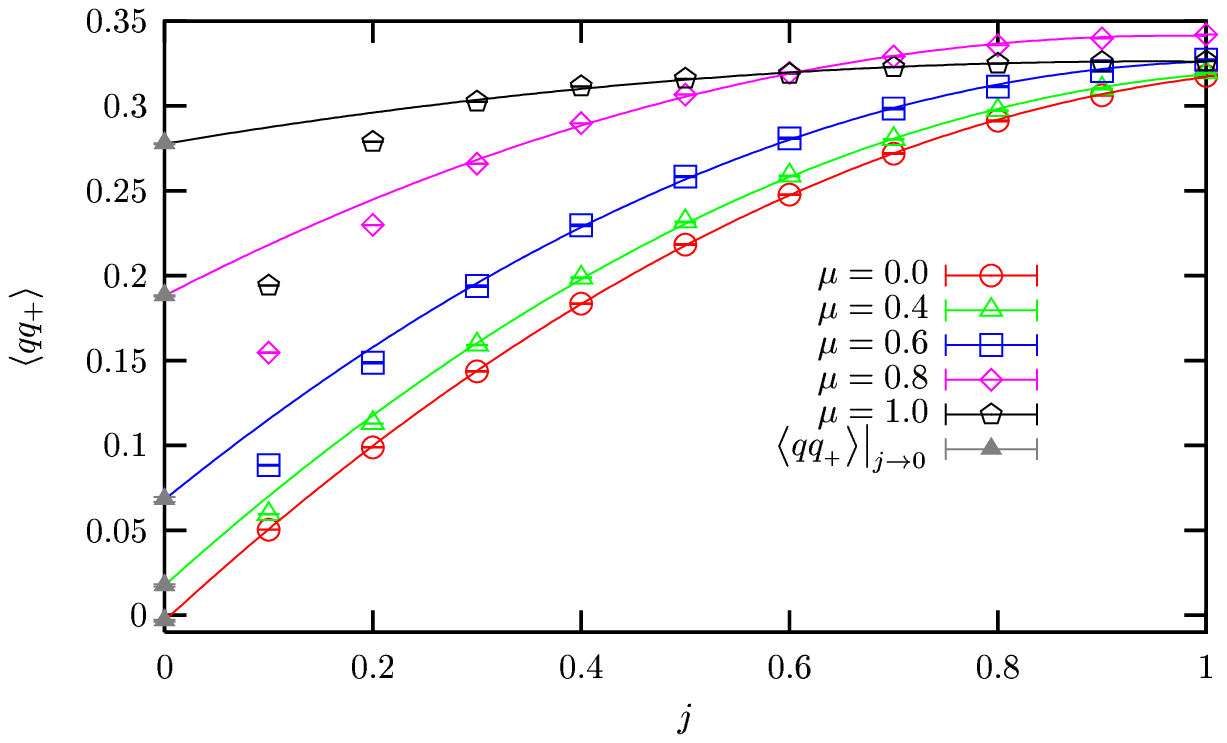}
\caption{$\left<q q_+\right>$ extrapolated to $j\to0$ for various
values of $\mu$.} 
\label{fig:qqvsj}
\end{figure}
For more direct evidence of diquark condensation we measure the order
parameter defined in (\ref{eq:qq}). Again, these data are extrapolated 
linearly to the limit $L_t^{-1}\rightarrow 0$ with the quality of the
fits being good. Figure~\ref{fig:qqvsj} shows the extrapolated
values of $\lt<q q_+\rt>$ plotted against $j$ for various values of $\mu$. 
Fitting a quadratic curve through the data with $j\ge0.3$, one can clearly
see that the high $\mu$, low $j$ data again disagree with the curve;
again ignoring these points, the data are extrapolated to $j\to0$. 
For $\mu=0$ we find no diquark condensation as one would expect, 
but as $\mu$ increases from zero, so does $\lt<q q_+\rt>$.
Together, the observations that $\lim_{j\to0}R=0$ and
$\lim_{j\to0}\langle q q_+\rangle\not=0$ 
support the existence of a BCS superfluid phase at high chemical potential. 

\begin{figure}[h]
\centering
\includegraphics[width=12cm]{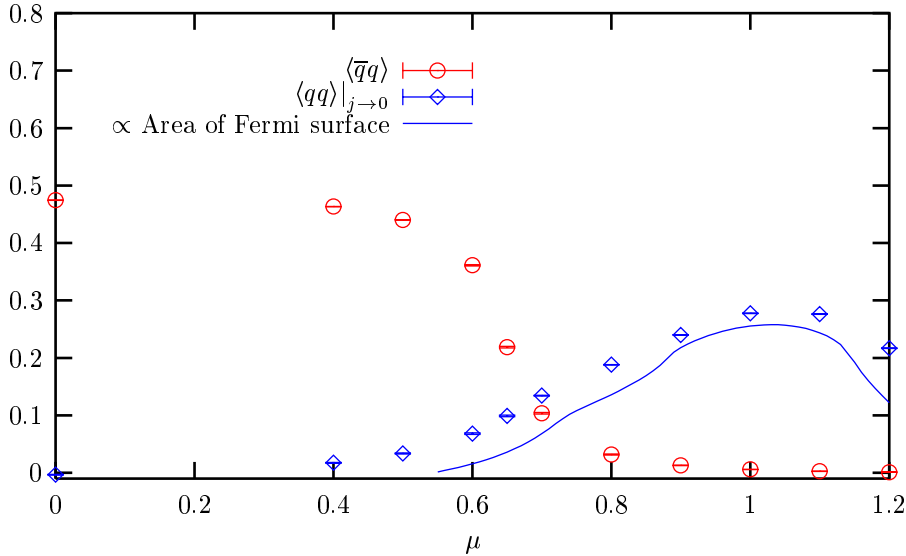}
\caption{A comparison between the diquark and chiral order parameters
as functions of $\mu$.} 
\label{fig:qqEofS}
\end{figure}
Finally, $\lt<q q_+\rt>$ is plotted as a function of $\mu$ in
Fig.~\ref{fig:qqEofS}, along with the previously presented result
for $\lt<\chibar\chi\rt>$.
 Although there is clearly a transition from a phase
with no diquark condensation to one in which the diquark condensate
has a magnitude similar to that of the vacuum chiral condensate,
this transition is far less pronounced than
in the chiral case. $\lt<q q_+\rt>$ increases approximately
as $\mu^2$, but eventually saturates as $\mu$ approaches 1.0 and even
decreases past $\mu\sim1.1$. This behaviour is 
directly related to the geometry of the Fermi surface for a system defined on a
cubic lattice, the 
area of which we have calculated in the large-$N_c$ free fermion limit from
(\ref{eq:Surface}) and have plotted as the solid curve. The method of
this calculation is sketched in Appendix~\ref{app:area}.
In the continuum, $\lt<q q_+\rt>$ should continue to rise as
$\mu^2$, such that the  curvature
$\partial^2\lt<q q_+\rt>/\partial\mu^2$ is positive, in contrast to the
behaviour observed in simulations of two color QCD, in which there is 
no Fermi surface and U(1)$_B$ breaking proceeds via
Bose-Einstein condensation~\cite{Aloisio:2000rb,Kogut:2003ju,Hands:2001ee}
leading to $\partial^2\langle q q_+\rangle/\partial\mu^2<0$.

The apparent weakness of the transition at intermediate $\mu$ is related to
the fact that at these chemical potentials, the value of $\lt.R\rt|_{j\to0}$
interpolates between the two extremes of 0 and 1. This is counterintuitive,
since it suggests a partially broken symmetry, even at $j=0$.
It may be that this is a side-effect of the chiral transition being a
crossover, since there is no sharp point at which a large Fermi surface
is created. 
It is also possible, of course, that this behaviour for intermediate
$\mu$  is merely an artifact of our poor control over the $j\to0$ extrapolation.


\section{The Quasiparticle Dispersion Relation}
\label{sec:disprel}

\subsection{Spectroscopy in the Fermionic Sector}
\label{ssec:spectrosc}

In this section we study the dynamics of the model's fermionic excitations,
which as in the original BCS
theory~\cite{Bardeen:1957mv} can be viewed as
quasiparticles  with energy $E$ relative to the system's Fermi energy $E_F$.
In a traditional Fermi liquid, these can be identified with {\em particle}
excitations above the Fermi surface, and {\em hole} excitations below,
both of which can have energies arbitrarily close to $E=0$. 
In a superfluid system, however, the particles and holes mix, and energies of the
lowest-lying excitations are separated from zero by a BCS gap
$\Delta$, which in analogy to the chiral mass gap in the
vacuum $\Sigma_0$, can be viewed as an effective order parameter for
the system. One advantage of this parameter is that unlike the
diquark condensate, $\Delta$ can be directly related to a macroscopic
thermodynamic property of the system, the critical temperature
$T_c$~\cite{Schmitt:2002sc}. In principle, being a spectral
quantity it is also measurable 
in a color superconducting phase in QCD, 
where according to Elitzur's theorem one cannot define
a local order parameter in a gauge invariant way~\cite{Elitzur:1975im}.

The propagation of the quasiparticles is described by the Gor'kov
propagator, defined in (\ref{eq:gorkov}), such that by analysing its
momentum dependence we can map out a quasiparticle dispersion
relation $E(k)$, i.e. the energy spectrum as a function of momentum, and hence 
measure $\Delta$. In particular, we have measured the time-slice form
of both ``normal'' and ``anomalous'' propagators
\bea
N(\vec k;t)=\sum_{\vec x}{\rm Re}\lt[\bar N_{11}(\vec0,0;\vec x,t)\rt]
e^{-i \vec k.\vec x};\\
A(\vec k;t)=\sum_{\vec x}{\rm I m}\lt[A_{12}(\vec0,0;\vec x,t)\rt]
e^{-i \vec k.\vec x},
\label{eq:timeslice}
\eea
on $L_x\times L_{y,z}^2\times L_t$
lattices with $L_x=96$, $L_{y,z}=12$. 
This choice of having one spatial direction much
longer than the others gives the system a large number of 
modes with which to sample $E(k)$,
whilst minimising the extra computational expense of running on a
larger volume. 
In particular, by setting $\vec k=(k,0,0)$ with $k=2\pi n/L_x$
$(n=0,1,2,\dots,L_s/4)$, the lattice fermions have 25 independent 
modes between 0 and $\pi/2$ in the $k_x$ direction. Since the study
presented in Sec.~\ref{sec:finitev} shows that the
diquark observables display little spatial dependence, 
there should be no detrimental effects from working with $L_x\gg
L_{y,z}$. 

Simulations were performed with $L_t=16$ and 20 at various
chemical potentials using the same values of the diquark sources
as in the previous section. $L_t=12$ data were
also generated, but these prove to have too few
time-slices over which to reliably fit the propagator, and are
therefore neglected in our analysis.
Again, approximately 500
equilibrated trajectories were generated per run, with measurement
taking place on every other configuration. Two additional simulations
were performed at $\mu=0.0$ and 0.8 with $L_t=24$, which after
equilibration took approximately $5\frac{1}{2}$ and $16$ CPU days
respectively to generate 400 trajectories on a 2.0GHz Intel Xeon
processor.

\begin{figure}[h]
\centering
\includegraphics[width=7.5cm]{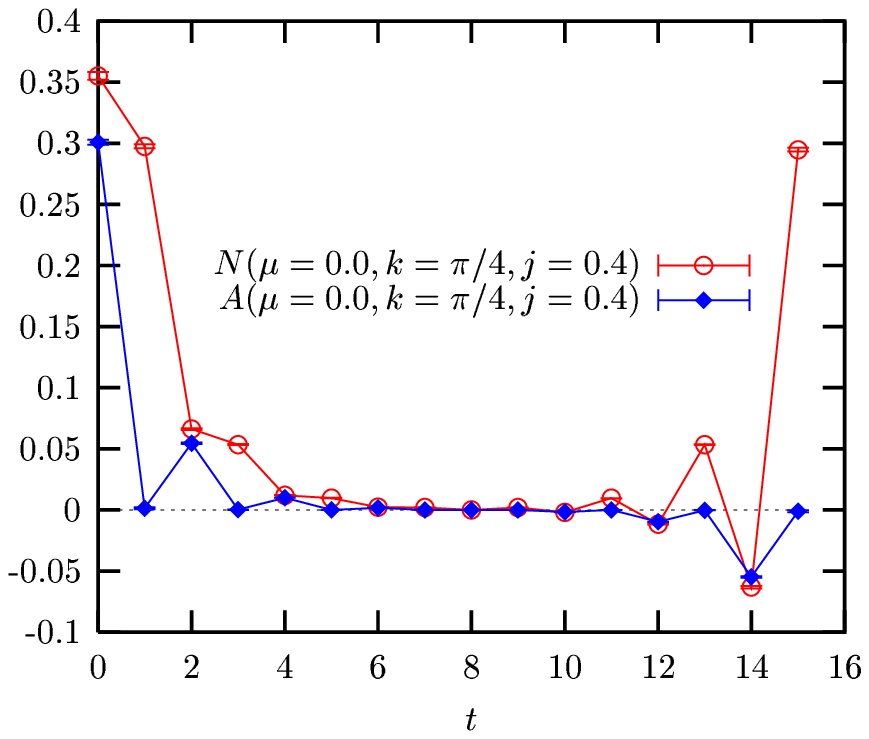}
\includegraphics[width=7.5cm]{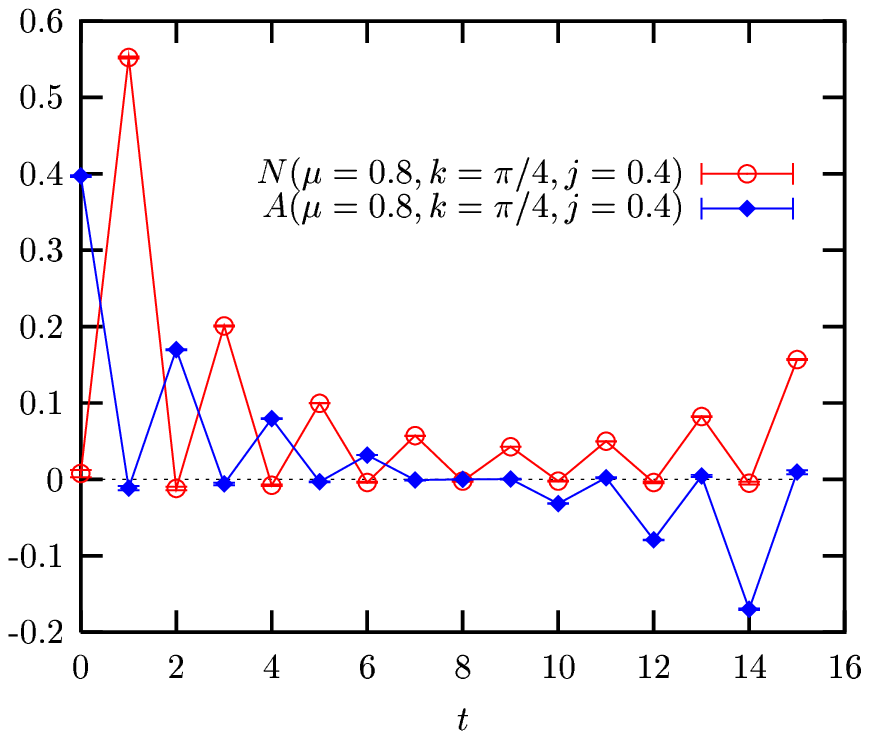}
\caption{Normal and anomalous propagators  measured on a $96\times12^2\times16$ lattice in both the chirally broken and restored phases.}  
\label{fig:NandA}
\end{figure}
Some example propagators in both the chirally broken and restored
phases are plotted in Fig.~\ref{fig:NandA}. At $\mu=0.0$ the normal
propagator is non-zero for all $t$, whilst at $\mu=0.8$ it
 approximates zero on even time-slices, which reflects the fact that
with manifest chiral symmetry 
the $N_{o o}$ and $N_{e e}$ components of the standard staggered fermion
propagator vanish. The anomalous propagator is zero on all odd
time-slices for all values of $\mu$.

\begin{figure}[ht]
\centering
\includegraphics[width=7.5cm]{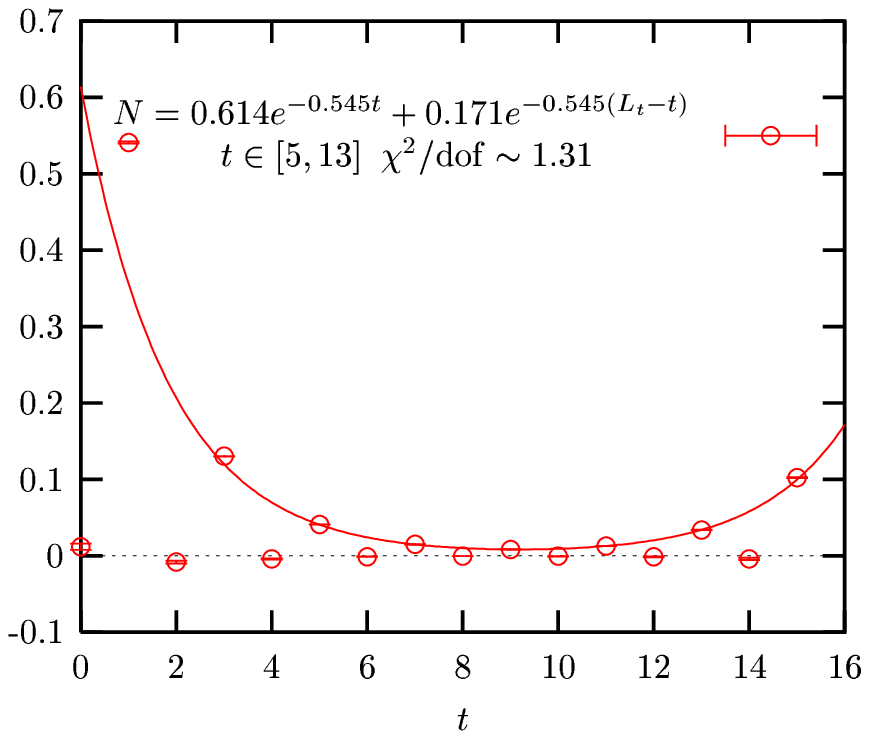}
\includegraphics[width=7.5cm]{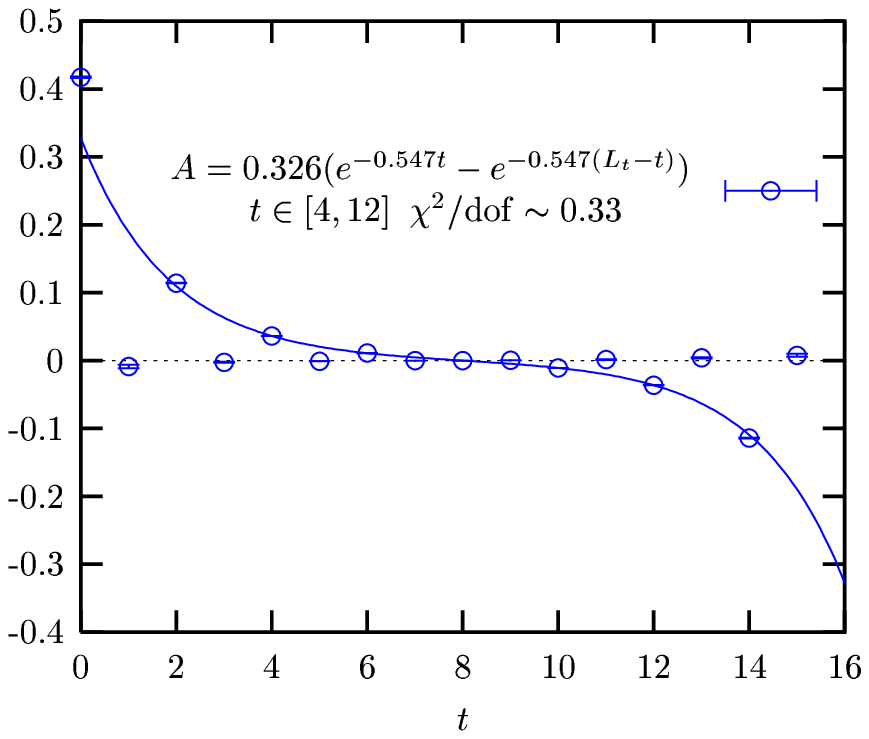}
\caption{Normal and anomalous propagators with $j=0.3$ and $k=\pi/4$
with their fitted curves  measured on a $96\times12^2\times16$ lattice
at $\mu=0.8$.}
\label{fig:NandAfitted}
\end{figure}

To map out the dispersion relation for each value of $\mu$, 
the energy is extracted by fitting the propagators to  
\be\ba{c l l}
N(k,t)=A e^{-E t}+Be^{-E(L_t-t)}&{\rm if}&t={\rm odd}\\
N(k,t)=0&{\rm if}&t={\rm even}
\label{eq:Nfit}
\ea\ee
and
\be\ba{c l l}
A(k,t)=C(e^{-E t}-e^{-E(L_t-t)})&{\rm if}&t={\rm even}\\
A(k,t)=0&{\rm if}&t={\rm odd},
\label{eq:Afit}
\ea\ee
where $A$, $B$ and $C$ are kept as free
parameters, as is the energy $E$. Although, as expected, the values of $E$
extracted from the two propagators are found to be consistent, we
choose to use those extracted from (\ref{eq:Afit}) to map out our
dispersion relation, since having
one less free parameter than (\ref{eq:Nfit}), the fits to this form
are found to be of a higher quality. Some example fits are
illustrated in Fig.~\ref{fig:NandAfitted}.

\begin{figure}[ht]
\centering
\begin{tabular}{l}
\includegraphics[width=6.5cm]{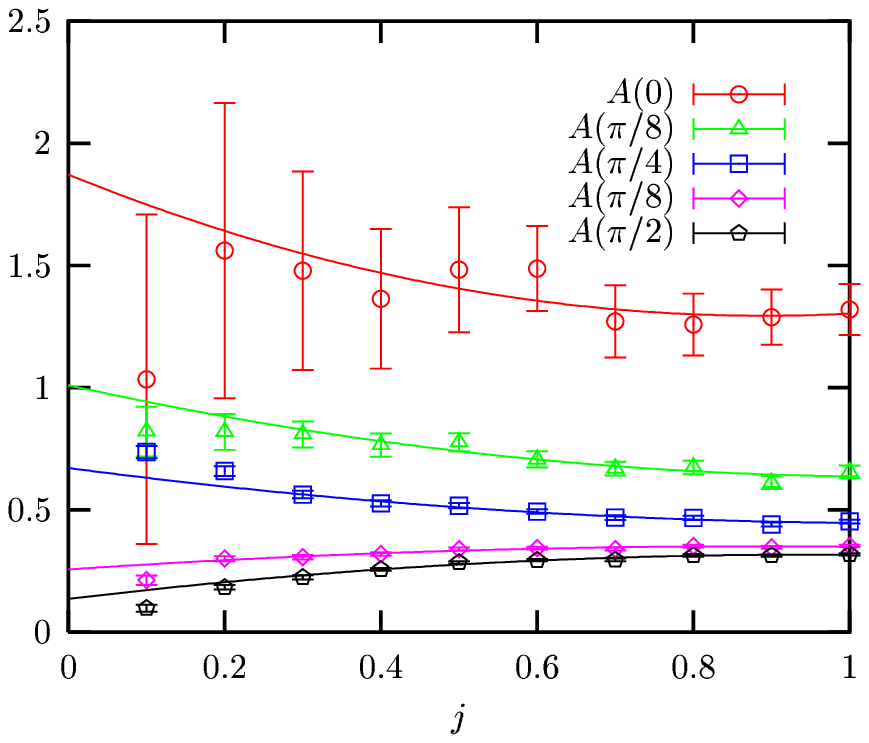}
\includegraphics[width=6.5cm]{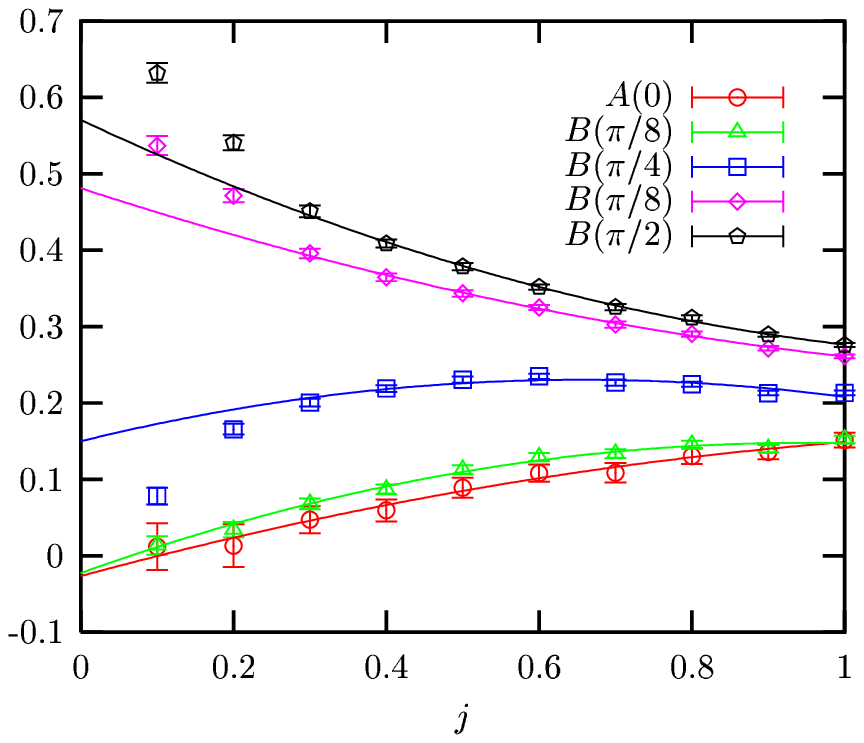}\\
\includegraphics[width=6.5cm]{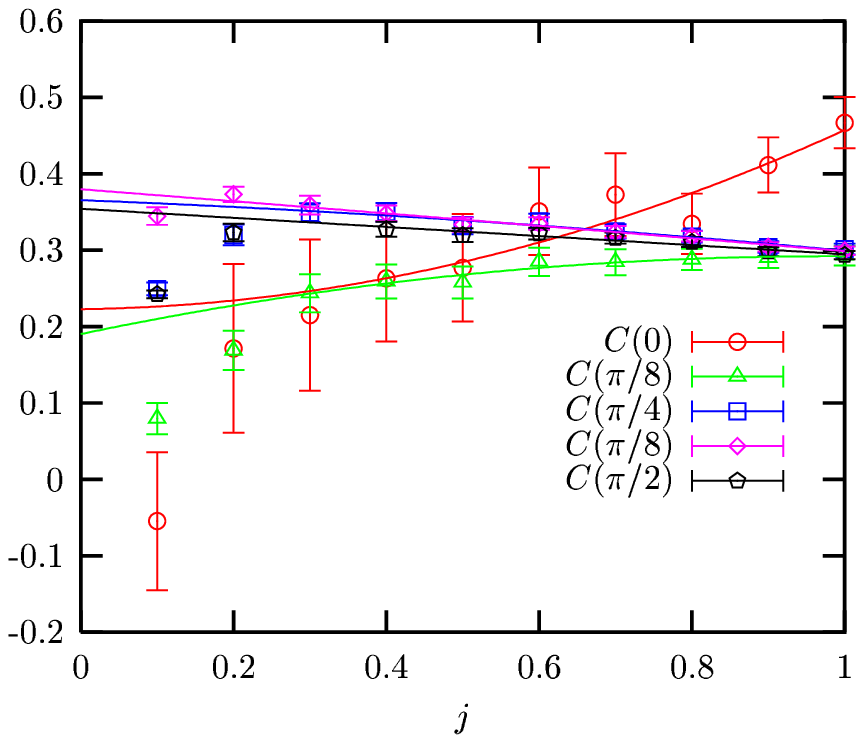}
\includegraphics[width=6.5cm]{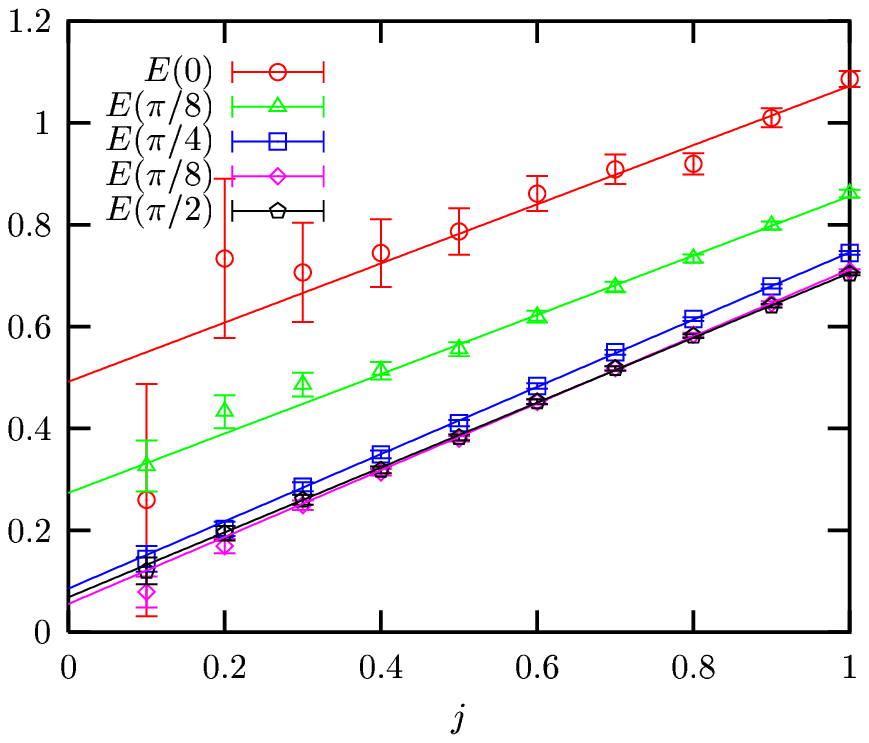}
\end{tabular}
\caption{Zero temperature propagator parameters at $\mu=0.8$ and
various values of $k$. The solid curves show extrapolation to $j\to0$.}  
\label{fig:AandEextrap}
\end{figure}
Figure~\ref{fig:AandEextrap} shows plots of the free parameters in
(\ref{eq:Nfit}) and  (\ref{eq:Afit}) at $\mu=0.8$,  extrapolated to $T\to0$;
in turn these are extrapolated to $j\to0$. Quadratic polynomial curves are
fitted to the coefficients $A(k)$, $B(k)$ and $C(k)$, whilst the
energy $E(k)$ is fitted linearly. As with the
extrapolations of $\lt<q q_+\rt>$ and $R$, these
appear to smoothly fit the data except
for those with low values of the diquark source $j$, for which the discrepancy
discussed in Sec.~\ref{ssec:qq} persists. Again, points with $j<0.3$ are
ignored for the purpose of the extrapolations; we rely on the
conclusions of Secs.~\ref{sec:finitev} and \ref{sec:Tneq0}
to justify the omission of these data from our analysis. 

\subsection{The Vacuum Dispersion Relation}
\label{ssec:vacdisp}

Before considering a lattice system with a Fermi surface, we
review the nature of the dispersion relation in the familiar
case of the vacuum. With $\mu=0$, time reversal is a good
symmetry of the lattice and the coefficients $A$ and $B$ become
identical such that 
(\ref{eq:Nfit}) reduces to its usual form $\lt|N(k,t)\rt|=A(e^{-E
t}+e^{-E(L_t-t)})$. This can be understood physically by noting that
the vacuum spectrum appears identical to both particles and antiparticles, and
hence to both the forward- and backward-moving parts of $N$. In agreement with
this, $A$ and $B$ are found to be equal, within errors, for all three
values of $L_t$.  
\begin{figure}[h]
\centering
\includegraphics[width=12cm]{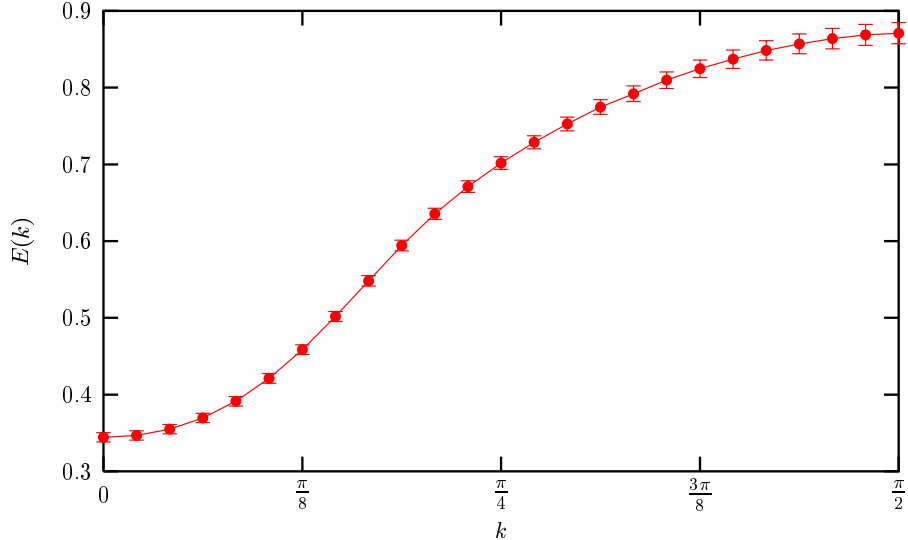}
\caption{The vacuum dispersion relation $E(k)$ extrapolated to $T\to0$
and then $j\to0$.} 
\label{fig:vacEvsk}
\end{figure}
Figure~\ref{fig:vacEvsk} illustrates the energies extracted from
$A(k,t)$, extrapolated to zero 
temperature through $L_t=16$, 20 and 24 and then to $j\to0$, which
results in the familiar lattice dispersion relation
$E(k,\mu=0)$~\cite{Boyd:1992uk}
\be
\sinh^2E=\alpha^2\sum_{i=1}^3\sin^2k_i+\Sigma_0^2,
\label{eq:disprel}
\ee
where $\alpha$ is a constant which allows for the renormalisation of the
speed of light by thermal effects;
in practice its fitted value is close to one as expected.
The energy gap at $k=0$ can be identified with the vacuum fermion mass,
from which we learn that $\Sigma_0=0.351(6)$. As $k$ is increased, the
dispersion relation is approximately quadratic for small $k/\Sigma_0$
(as expected for a non-relativistic particle), 
until discretisation effects dominate its form and the
periodicity of (\ref{eq:disprel}) causes it to level off as $k\to\pi/2$.

\subsection{Measurement of the Gap}
\label{ssec:gap}

\begin{figure}[h]
\centering
\includegraphics[width=12cm]{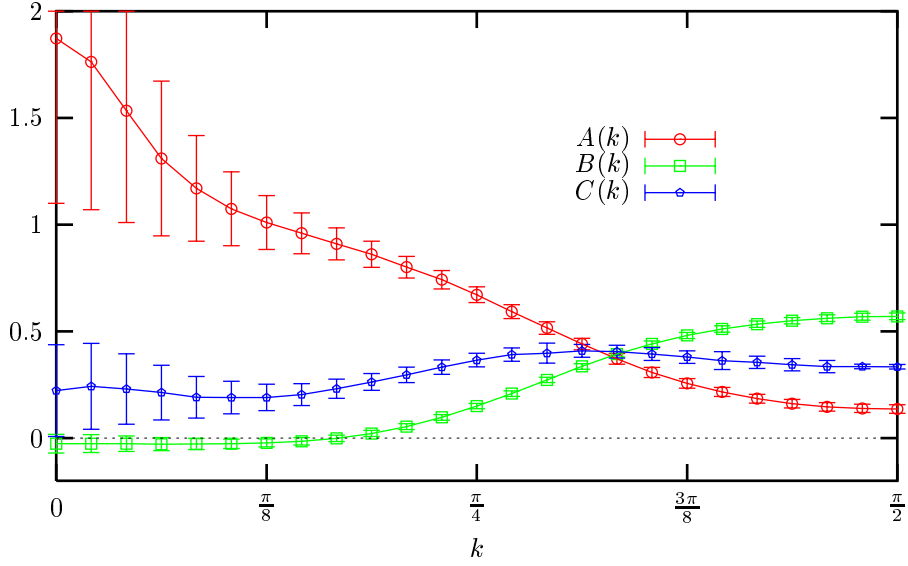}
\caption{The propagator coefficients $A$, $B$ and $C$ for $\mu=0.8$
extrapolated to $T\to0$ and then $j\to0$.}
\label{fig:ABCvsk}
\end{figure}

In this section we study the dispersion relation at $\mu=0.8$, with the
aim of observing the BCS gap $\Delta$. First, however, we study the
momentum dependence of the other free parameters fitted from the forms
(\ref{eq:Nfit}) and (\ref{eq:Afit}). Figure~\ref{fig:ABCvsk}
illustrates the values of $A$, $B$ and $C$ extrapolated first to $T\to0$
and then to $j\to0$, and plotted as functions of
momentum $k$.

The coefficients $A(k)$ and $B(k)$ represent the amplitudes of 
forward- and backward-moving spin-$\frac12$ propagation, which due to
our choice to study the antiparticle propagator
$N\equiv\overline{N}_{11}={\cal G}_{3 1}\sim\lt<\overline q(x) q(y)\rt>$,
correspond to hole and particle excitations respectively in the
limit that $j\to0$. For small momenta, corresponding to excitations deep
within the Fermi sea, propagation is dominated by hole degrees of
freedom. As the Fermi momentum is approached, particles are
easier to excite and become the dominant contribution as
$k\to\frac\pi2$. To degrees of freedom with $k=k_F$, as in the vacuum the
background appears the same to both particles and holes such that
$A(k_F)=B(k_F)$. For this reason, in analogy with the coefficients of
filled and unfilled states in the original BCS
theory~\cite{Bardeen:1957mv}, the intercept of the curves of $A(k)$
and $B(k)$ defines the Fermi momentum for the
interacting theory in the $T\to0$ limit.
In the anomalous sector,
the coefficient $C(k)$ is approximately zero deep within the Fermi
sea, but becomes non-zero (even in the limit $j\to0$) in a broad peak
about the Fermi momentum, which is a sign of particle-hole mixing
in this region. This is in contrast with similar measurements in 
NJL$_{2+1}$~\cite{Hands:2001aq}.

\begin{figure}[h]
\centering
\includegraphics[height=6cm]{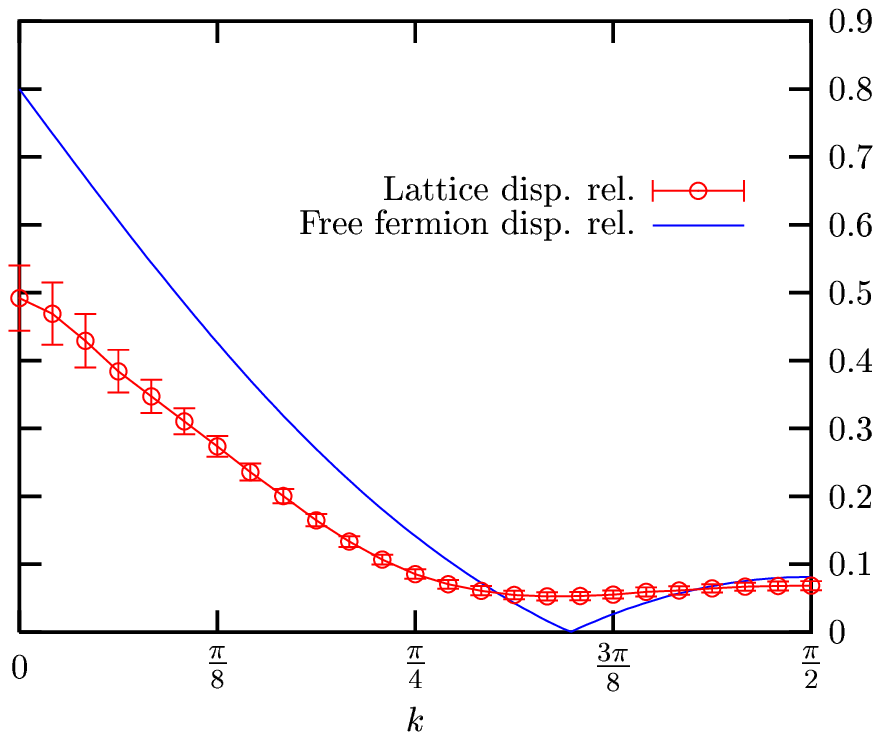}
\includegraphics[height=6cm]{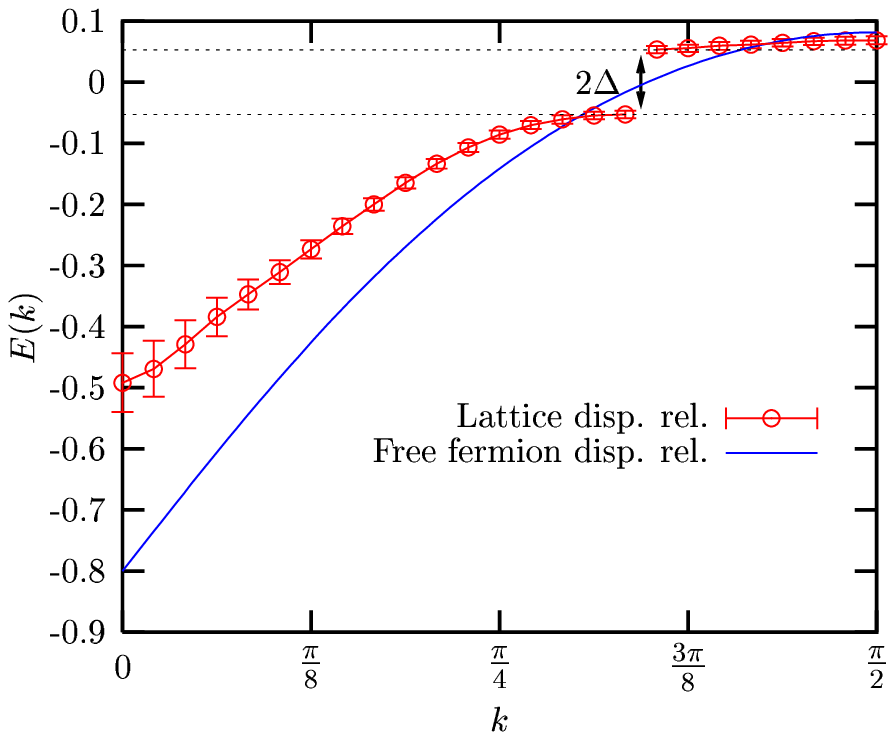}
\caption{The lattice dispersion relation and typical free fermion
dispersion relation at $\mu=0.8$. In the right-hand panel the hole
branch is plotted as negative.}
\label{fig:mu_0.8_Evsk}
\end{figure}
The left-hand panel of Fig.~\ref{fig:mu_0.8_Evsk} 
illustrates the  lattice dispersion relation $E(k)$ at $\mu=0.8$ extracted from
(\ref{eq:Afit}) and extrapolated first to $T\to0$ and then $j\to0$ (points),
compared with that of free massless fermions on the lattice, parametrised by
\be
E(k)=-\mu+\sinh^{-1}(\sin k).
\ee
For this free dispersion relation, there are two distinct branches, one
for hole excitations below the Fermi surface for which $E$ reduces with
$k$, and one for particle excitations for which $E$ increases with
$k$. We should note at this point that the crossover between these
regimes, $k_F=\sin^{-1}(\sinh0.8)\sim0.348\pi$, is consistent with the
intercept of $A(k)$ and $B(k)$ in Fig.~\ref{fig:ABCvsk} to within the
precision allowed by the momentum resolution. By contrast the
lattice data display no evidence for two distinct branches,
which is another signal of particle-hole mixing. More importantly, at
no point do the lattice data pass through zero, 
but between $\pi/3\le k\le17\pi/48$ (again consistent with the
free-field $k_F$), they have a minimum of $E(k)=0.053(6)$; this is the BCS
gap $\Delta$. Comparing this to our measurement of the vacuum fermion
mass in Sec.~\ref{ssec:vacdisp}, we find the ratio
\be
\frac{\Delta(\mu=0.8)}{\Sigma_0}=0.15(2),
\ee
which assuming a fermion mass of 400${\rm M e V}$ implies that
$\Delta(\mu=0.8)\sim60{\rm M e V}$, consistent with the analytic predictions
of~\cite{Berges:1998rc,Nardulli:2002ma}.

This may be viewed more graphically in the right-hand panel of
Fig.~\ref{fig:mu_0.8_Evsk}, where excitations below $E_F$ have been
plotted as negative. This makes the free fermion dispersion relation a
smooth curve that passes continuously through $E=0$ at the Fermi
momentum, and is similar in nature to that of lattice four-Fermi models with no BCS
gap~\cite{Hands:2001aq,Hands:2003dh}. For our data, however,
this plot introduces a discontinuity of $2\Delta$ which looks exactly like
that of a traditional BCS superfluid in the continuum.

\subsection{$\mu$ Dependence of the Gap}
\label{ssec:mudep}

Having systematically investigated the dispersion relation and
extracted the gap at one value of chemical potential, it would be
illuminating to repeat the analysis of the previous section for a range
of chemical potentials in the BCS phase. 
However, generating data with $L_t=24$ in the
chirally restored phase is a CPU intensive task, taking $O
  $(20) CPU
days on a fast desktop PC for each value of $\mu$.
The reason this is so much more expensive than in the chirally broken phase
is that the rate of convergence of the conjugate
gradient subroutine used to invert $M^\dagger M$ in the generation of our
$\{\Phi\}$ configurations is related to the magnitude of the
diagonal components of $M$, which are in turn proportional to the
constituent quark mass $\Sigma(\mu>\mu_c)\simeq0$.

For this reason we utilise the data generated with $L_t=16$ and
20 and approximate the $T\to0$ limit by extrapolating through these
data and assigning a conservative estimate for the error; the $j\to0$
extrapolations may then be carried out as before. Although of little
statistical significance, it should be noted that the dispersion
relation at $\mu=0.8$ extracted using this method is consistent, for
all values of $k$, with that extracted using the full statistical
treatment of the previous section.

\begin{figure}[h]
\centering
\includegraphics[width=7cm]{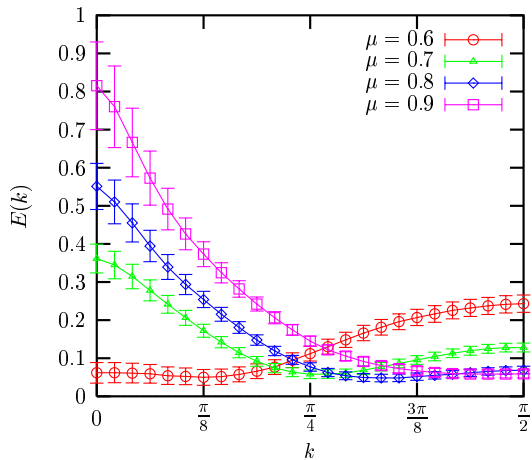}
\caption{Lattice dispersion relations at various values of $\mu$.}
\label{fig:Evsk}
\end{figure}
\begin{figure}[h]
\centering \includegraphics[width=10cm]{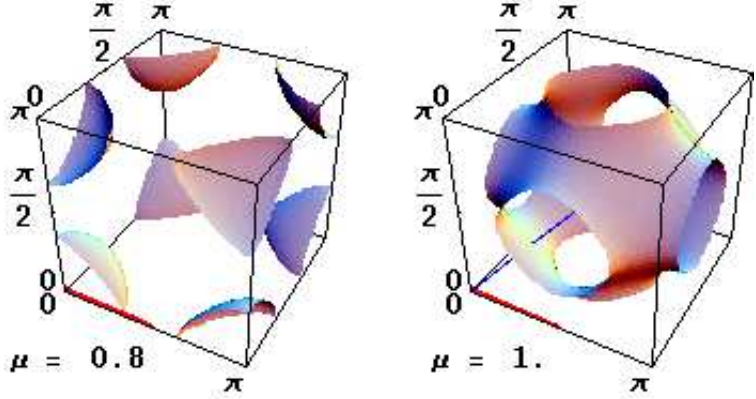}
\caption{The zero temperature, infinite volume Fermi surface at two
values of chemical potential.}
\label{fig:surface}
\end{figure}

Figure~\ref{fig:Evsk} illustrates a selection of these dispersion
relations for various chemical potentials. An interesting point to
note is that unlike the diquark condensate, there is an upper bound of
$\mu\sim\sinh^{-1}(1.0)\sim0.88,$ above which we cannot extract an
estimate of the gap using the method outlined in
Sec.~\ref{ssec:spectrosc}. This can be understood by considering
the nature of the lattice Fermi surface for free fermions in the
infinite volume limit, parametrised by (\ref{eq:Surface})
and plotted at two values of $\mu$ 
in the large-$N_c$ limit in Fig.~\ref{fig:surface}.
For ${k_F}_i\ll\pi/2$, corresponding to small values of $\mu$, 
the small-angle identities imply that the surface is
approximately spherical and the momenta sampled herein, emphasised on
the $k_x$ axis in Fig.~\ref{fig:Evsk}, are sufficient to probe either
side of the surface and detect the presence of a gap. For
$\mu>\sinh^{-1}(\sin{\pi/2})$, however, the reduced rotational
invariance of the lattice dominates, and (\ref{eq:Surface}) has no
real solution with $\vec k=(0\le k_x\le\frac\pi2, 0, 0)$.  It is
for this reason that the curve at $\mu=0.9$ in Fig.~\ref{fig:Evsk}
shows no minimum and we cannot extract a value of $\Delta$. Whilst
there is almost certainly a gap present, as the maximum of $\lt<q
q_+\rt>$ lies between $1.0\le\mu\le1.1$, to detect its presence via the
dispersion relation one is required to sample momenta along 
e.g. the more complicated diagonal path $\vec k=(k,k,k)$ with $0\le
k\le\pi/2$, illustrated in the right-hand panel of
Fig.~\ref{fig:surface}. Because of the large number of spatial modes
required to sample this path with sufficient resolution, such a
study is computationally beyond our current capability.

\begin{figure}[h]
\centering
\includegraphics[width=7cm]{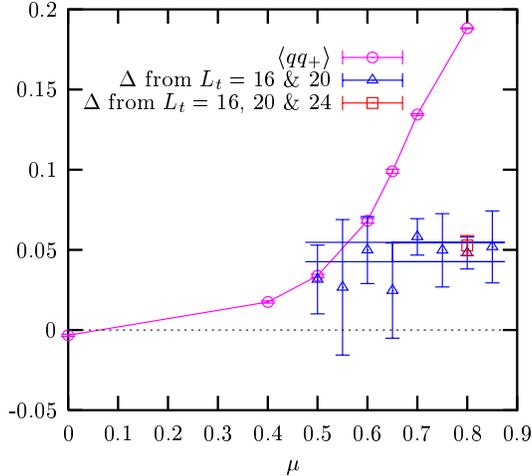}
\caption{Estimates for the gap $\Delta$ compared with $\left<q
q_+\right>$ for a range of $\mu$.} 
%
%
\label{fig:Evsmu}
\end{figure}

Finally, $\Delta$ is plotted as a function of $\mu$ in
Fig.~\ref{fig:Evsmu} and compared with the value of the diquark
condensate. Although the error bars on our estimated values are fairly
large, we see little evidence of any $\mu$ dependence once the gap
becomes non-zero. In fact, a least-squares fit to $\Delta={\rm
constant}$, denoted by the horizontal bar, has a chi-squared value of only
0.33 per degree of freedom. In combination with Fig.~\ref{fig:qqEofS}, 
Fig.~\ref{fig:Evsmu} provides qualitative
support for a simple-minded picture in which only diquark pairs within
a shell encasing the Fermi surface of thickness $2\Delta$, 
independent of $\mu$, participate in the
pairing, resulting in a condensate $\langle q q_+\rangle\propto\Delta\mu^2$.


\section{Finite Spatial Volume Effects}
\label{sec:finitev}

The conclusions of Secs.~\ref{sec:phase} and \ref{sec:disprel},
that the high-$\mu$ phase is one with both $\lt<q q_+\rt>\neq0$ and
$\Delta\neq0$, both rely on the discarding of data with $j<0.3$, since
results in the diquark sector with these small diquark sources
disagree with the trends at higher $j$. In order to be able to trust
our $j\to0$ extrapolations it is necessary, therefore, to justify this
omission, especially since it is the data in the region of this limit
that have been discarded. 

We have previously argued that the discrepancy at low-$j$ could be due 
to finite size effects~\cite{Hands:2002mr}, since whilst our
simulations were performed on lattices with $L_s\alt20$, variational
studies of the $N_f=2$, $N_c=3$ continuum NJL model at zero
temperature in a finite spatial volume show that
with no diquark source, a spatial extent of 7fm ($\sim25$ lattice spacings) is
required before the model approximates its infinite volume
limit~\cite{Amore:2001uf}. We have argued further that the source of
these finite size effects is due not to the realisation of an exact
Goldstone-mode, but to the difficulty of representing a thin shell of 
states about the Fermi surface which contribute to diquark
condensation on a discrete momentum lattice. Whilst this is supported
in part by the finite size study presented in~\cite{Hands:2002mr},
in which $\lt<q q_+\rt>$ displays a non-monotonic dependence on $L_s$, 
it should be noted that the magnitudes of these fluctuations are less
than 1\% for $L_s\ge8$ and all values of $j$, much smaller than
the approximate $30\%$ suppression of $\lt<q q_+\rt>$ at $j=0.1$ seen in
Fig.~\ref{fig:qqvsj}. In
fact, the diquark condensate shows little notable $L_s$ dependence
even prior to extrapolation to $T\to0$. 
\begin{figure}[h]
\centering
\includegraphics[width=12cm]{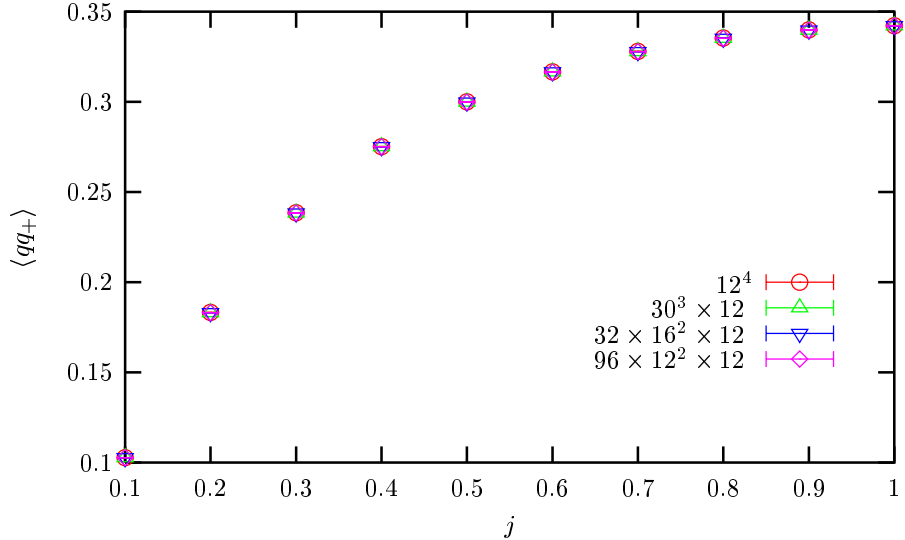}
\caption{Diquark condensate at $\mu=0.8$ on $L_x\times L_{y,z}^2\times
L_t$ lattices with $L_t=12$ and various $L_{x,y,z}$.}
\label{fig:qq_lt_12}
\end{figure}
In Fig.~\ref{fig:qq_lt_12}, $\lt<q q_+\rt>$ is plotted as a function
of $j$ at $\mu=0.8$  with $L_t=12$ and various spatial extents, including the
results of a large simulation with $L_s=30$. This shows that the diquark
condensate displays no significant size dependence at any $j$. The
same lack of any significant spatial size dependence is found for both
$\lt<q q_+\rt>$ with $L_t=16$ and 20 and for $E(k)$ with $L_t=16$ and
20 within the allowed momentum resolution. 
The fact that these observables remain
unchanged at low $j$, even  when changing the
spatial volume by up to a factor of $2.5^3\sim16$, makes it hard to believe
our previous suggestion that the suppression at $j<0.3$ is due to
finite size effects. Instead, we propose an alternative suggestion in
the following section.


\section{Non-zero Temperature}
\label{sec:Tneq0}

In previous reports of this work~\cite{Walters:2003it,Walters:2003nn},
we have suggested that an obvious extension would be to carry out
simulations at non-zero temperature, with the aim of measuring the
critical temperature of the superfluid phase, $T_c$. The
non-relativistic BCS theory predicts the relation
between this parameter and the magnitude of the gap to
be~\cite{Bardeen:1957mv}
\be
\frac{\Delta}{T_c}=1.764.
\label{eq:Tcratio}
\ee
Furthermore, it has been shown that this relation holds for
relativistic color superconducting systems in weakly coupled QCD with
two flavors~\cite{Schmitt:2002sc}. Although such weak coupling
predictions may be trusted only at asymptotically high densities,
\naively applying (\ref{eq:Tcratio}) to our measurement of 
$\Delta(\mu=0.8)=0.053(6)$ suggests that $T_c\sim0.03a^{-1}$,
such that at this chemical potential and in the limit $j\to0$,
one should observe a superfluid phase only when the temporal extent is
greater than about 35 lattice spacings. The fact that we observe a BCS
phase, even though our simulations were performed on lattices with
temporal extents much smaller than this relies on our performing measurements
with $j\neq0$ and then extrapolating in the correct manner. 
Setting $j\neq0$ has the effect of making condensation more
favourable, which suggests that at fixed $j$ there
could be a crossover at some pseudo-critical temperature, $T_{c}(j)$,
separating a region where diquark condensation is suppressed by thermal
fluctuations and one in which it is not. One would expect the
effect of increasing the source would be to increase $T_c(j)$ such
that diquark condensation can be observed on lattices with smaller
temporal extents. If one can successfully extrapolate to zero
temperature first, a $j\to0$ extrapolation should then be possible.

\begin{figure}[h]
\centering
\includegraphics[width=12cm]{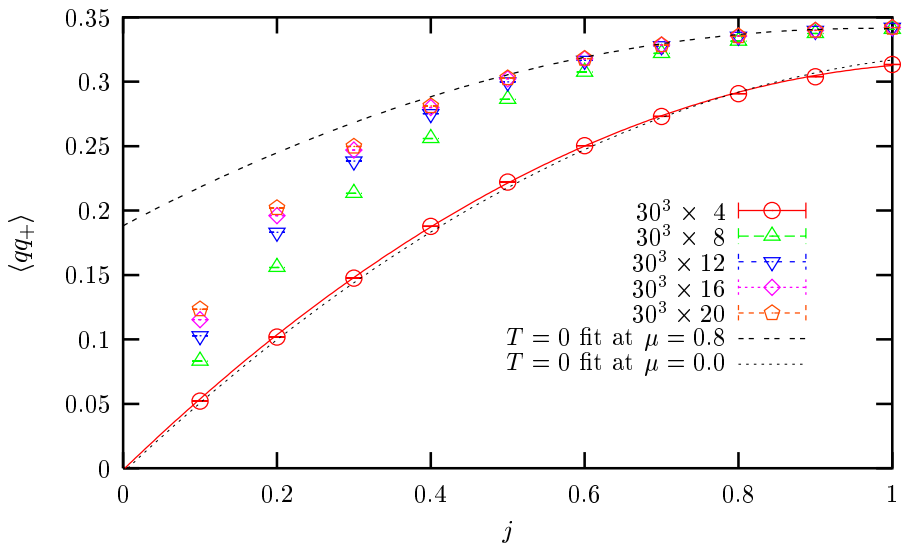}
\caption{Diquark condensate at $\mu=0.8$ on various lattices at
non-zero temperature.} 
\label{fig:Tneq0}
\end{figure}

\begin{figure}[h]
\centering
\includegraphics[width=12cm]{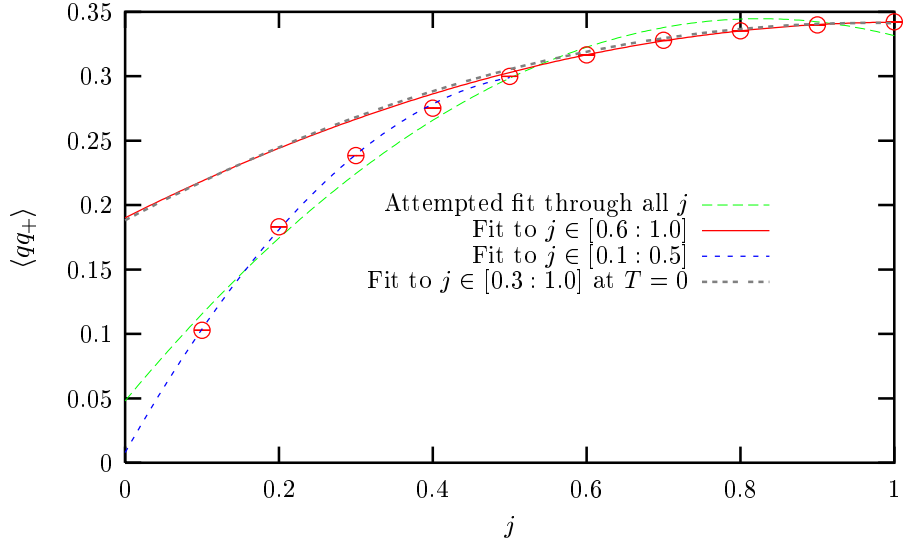}
\caption{Diquark condensate at $\mu=0.8$ an a $30^3\times12$ lattice.} 
\label{fig:12x30to3}
\end{figure}

This causes a problem if one wishes to determine the value of the
condensate at a particular chemical potential and non-zero
temperature, since it is possible that at temperatures close to $T_c$,
$T_c(j)$ crosses the
temperature of the lattice over the range of $j$
studied. Figure~\ref{fig:Tneq0} 
illustrates the diquark condensate measured at $\mu=0.8$ on lattices
with various temporal extents corresponding to various non-zero
temperatures, as well as the ``$T=0$'' curves for $\mu=0.0$ and 0.8 
from $12^4$, $16^4$ and $20^4$ lattices, as plotted in Fig.~\ref{fig:qqvsj}.
The $L_t=4$ results lie well below the $T=0$ curve for all values of
$j$, suggesting that the temperature is high enough to suppress
condensation for the entire range of $j$. A quadratic extrapolation through these points
is very similar to the zero temperature vacuum fit from
Fig.~\ref{fig:qqvsj} and is consistent with 
$\lt.\lt<q q\rt>\rt|_{j\to0}=0$. As the temperature of the lattice 
is decreased,
however, the data at higher $j$ are no longer suppressed and an
extrapolation through all $j$ is no longer possible. This can 
be seen more clearly in Fig.~\ref{fig:12x30to3}, where we focus on the data
measured on a $30^3\times12$ lattice.
Whilst an attempted fit through all values of $j$ appears of poor
quality, by choosing a suitable point to separate the low- and high-$j$
data one can fit the two regions successfully. 

Whilst this implies that establishing the critical temperature of the
superfluid phase is not as simple as we initially believed, it does
provide an alternative explanation of the suppression in $\lt<q
q_+\rt>$ at $T=0$ for $j<0.3$. The fact that the curve fitted through
the $L_t=12$ data over $j\in[0.6:1.0]$ agrees with that of data
extrapolated to $T\to0$ and fitted over $j\in[0.3,1.0]$
suggests that these curves represent not the
correct infinite volume limit, as previously suggested, 
but the correct zero temperature
limit of the $j\to0$ extrapolation. Whilst a linear $T\to0$ extrapolation is
sufficient to reach this curve for $0.3\le j\le0.5$, the data at
$j<0.3$ must be suppressed too much for such an extrapolation to be
sufficient, i.e. the temperature of the lattice must be too high compared with
$T_c(j<0.3)$. This justifies in retrospect the discarding of the low $j$ data
in the extrapolations of Secs.~\ref{sec:phase} and \ref{sec:disprel}, which form
the basis for the claimed superfluid ground state.


\section{Non-zero Isospin Chemical Potential}
\label{sec:mu_Ineq0}

In previous sections, we have demonstrated numerically that the
ground-state of our two flavor model at non-zero $\mu$ and low $T$ is
that of a BCS gapped lattice superfluid. The pairing mechanism in this
system, between quarks of opposite momenta and isospin, and thus 
analogous to 2SC condensation between $u$ and $d$ quarks in QCD, is
particularly energetically favourable. If the system were
non-interacting, it would cost no energy to create a pair of quarks 
at the common Fermi surface of the degenerate flavors, such that when
the attractive interaction is restored, the system is expected always
to be unstable to diquark condensation~\cite{Alford:2000ze}.

In nature, however, the Fermi momenta $k_F^u$ and $k_F^d$ for up and
down quarks respectively are expected to differ. A
simplistic argument outlined in~\cite{Alford:1999pb}, simplified
still further here to describe a two flavor system, suggests that in compact
stellar matter $k_F^u$ should be less than $k_F^d$. For massless
non-interacting matter with baryon chemical potential $\mu_B=400{\rm M
e V}$, an electron chemical potential $\mu_e=89{\rm M e V}$ is
required to enforce both charge neutrality and chemical equilibrium
under weak interactions.
Together, these two conditions determine all the chemical potentials
and Fermi momenta: 
\be
\ba{c}
k_F^u=\mu_u=\mu_B-\frac{1}{2}\mu_e=355.5{\rm MeV},\\
k_F^d=\mu_d=\mu_B+\frac{1}{2}\mu_e=444.5{\rm MeV},\\
k_F^e=\mu_e=89{\rm MeV}.
\label{eq:muIpred}
\ea
\ee
Here we use the term baryon
chemical potential in the context of the NJL model, where baryons 
are identified with quarks and $\mu_B\equiv\frac12(\mu_u+\mu_d)$.  
The effect of separating the free-particle Fermi surfaces of pairing
quarks should be to make the superconducting phase less energetically
favourable, and should prove a good method to investigate the
stability of the superfluid phase.

Such a study was applied to the 2SC color superconducting phase
in a mean field four-Fermi model in~\cite{Bedaque:1999nu}. Similar to
results for superconductors in the presence of a magnetic
field~\cite{Clogsten,Chandrasekhar}, when the free field Fermi
surfaces are separated by only a small amount the ground-state of
interacting matter remains superconducting with degenerate Fermi
surfaces for the pairing 
partners. At some critical free fermion separation, however, the system is
found to go through a first order transition to a gapless Fermi liquid
with two separate surfaces. Unlike normal superconductors, however, the size
of the gap increases slightly under small flavor asymmetries, 
an effect attributed to the model's color structure extracted from one gluon
exchange in QCD. This analysis has also been extended further to
include systems in which the Fermi surfaces $k_F^u$ and $k_F^d$ are
separated not only by an isospin chemical potential, but a fixed
momentum $\vec q$~\cite{Alford:2000ze}. In such a system, when 
$\lt|k_F^u-k_F^d\rt|$ and
$\vec q$ are sufficiently large that the Fermi surfaces cross, Pauli
blocking implies that 2SC becomes unstable with respect to a state in
which diquark condensation occurs only at a ring of
states close to the intercept of the surfaces. The
state has both broken translational and rotational invariance in
which the diquark pairs have non-zero total momentum; in analogy with
such phases in electron superconductivity this is known as
the LOFF phase~\cite{Larkin,Fulde}.

In the lattice NJL model, the pairing quarks of opposite isospin are
represented by the two components of the staggered fermion field
\be
\chi_x\equiv\lt(\begin{array}{c}\chi_1(x)\\\chi_2(x)\end{array}\rt)
=\lt(\begin{array}{c}u(x)\\d(x)\end{array}\rt), 
\ee
hereon referred to as ``up'' and ``down'' quark
flavors. The Fermi surfaces of the pairing partners can be separated  
by directly  allocating them different chemical
potentials, $\mu_u$ and $\mu_d$, equivalent to
having simultaneously non-zero baryon chemical potential
$\mu_B=\frac{1}{2}\lt(\mu_u+\mu_d\rt)$ and isospin chemical potential 
$\mu_I=\frac{1}{2}\lt(\mu_u-\mu_d\rt)$.
Although this definition implies 
that $\mu_u>\mu_d$, which is contrary to the conclusions of the 
argument outlined above,
this notation has been chosen to be consistent with the analytic
studies  of~\cite{Toublan:2003tt} and~\cite{Frank:2003ve} (since the
NJL model does not include weak interactions and 
is therefore isospin invariant, the labels 
$u$ and $d$ are 
interchangeable). In the physical context of compact stars, the two scales
should be ordered $\mu_B\gg\mu_I$, since the simple argument leading
to (\ref{eq:muIpred}) predicts that
$\frac{1}{2}\lt|\mu_u-\mu_d\rt|\sim0.1\mu_B$.   

With this introduction, the fermion kinetic operator $M$, defined
previously in (\ref{eq:M}) becomes
\bea
M^{p q}_{x y} & = & \frac{1}{2}
\lt(e^{\mu_B}(e^{\tau_3\mu_I})^{p q}\delta_{y x+\hat{0}}
-e^{-\mu_B}(e^{-\tau_3\mu_I})^{p q}\delta_{y x-\hat{0}}\rt)\nonumber \\&+&
\frac{1}{2}\delta^{p q}
\lt[\sum_{\nu=1}^3 \eta_\nu(x)\lt(\delta_{y x+\hat{\nu}} -\delta_{y
x-\hat{\nu}}\rt) +2m\delta_{x y}\rt] 
\label{eq:Miso}\\  & + &
\frac{1}{16}\delta_{x y}\sum_{\lt<\tilde{x},x\rt>}
\lt(\sigma(\tilde{x})\delta^{p q} + i\epsilon(x)\vec{\pi}
(\tilde{x}).\vec{\tau}^{p q}\rt)\nonumber .
\eea
Unfortunately, this means that the proof that $\det M$ is real and
positive presented in~\cite{Hands:2001aq} is no longer
valid, which can be seen be noting that e.g.
$\tau_2(e^{\tau_3\mu_I})\tau_2=e^{-\tau_3\mu_I}\ne(e^{\tau_3\mu_I})^*$,
and the identity $\tau_2M\tau_2=M^*$ no longer holds.
Although this would not cause the simulation to fail, since we use
$\det M^\dagger M$ as our fermionic measure, the fact that this choice
is the sole reason the algorithm would work implies that 
non-trivial interactions between $\chi$ and $\zeta$
quarks will be introduced 
which could cause the argument of~\cite{Barbour:1999mc} to
break down. Instead, we choose to simulate 
in the quenched isospin limit in which $\det M^\dagger M$
is calculated with $\mu_I=0$ in the HMC update of the bosonic fields, whilst
(\ref{eq:Miso}) is used in ${\cal A}$ during the measurement of the
observables.  

Before we discuss setting $\mu_I\neq0$ in the superfluid phase,
let us examine the effect this introduction has on the chiral phase
transition. An analytic study of the NJL model has shown that
introducing a small $\mu_I$ causes the chiral phase transition 
to split into two, one transition for the condensate of each quark
flavor~\cite{Toublan:2003tt}. 
\begin{figure}[h]
\centering
\begin{tabular}{c c}
\includegraphics[width=7cm]{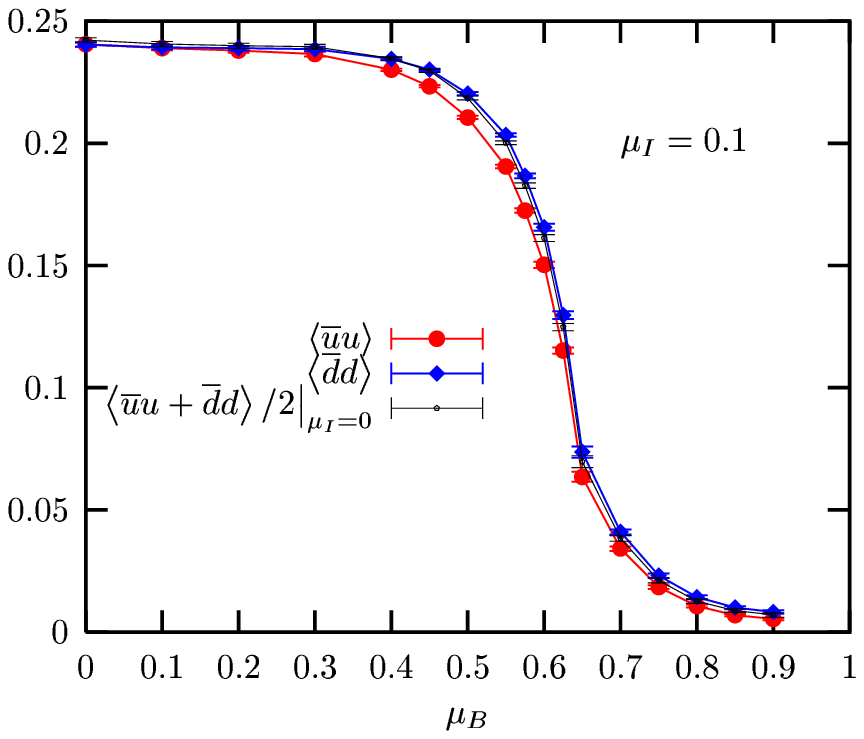}&
\includegraphics[width=7cm]{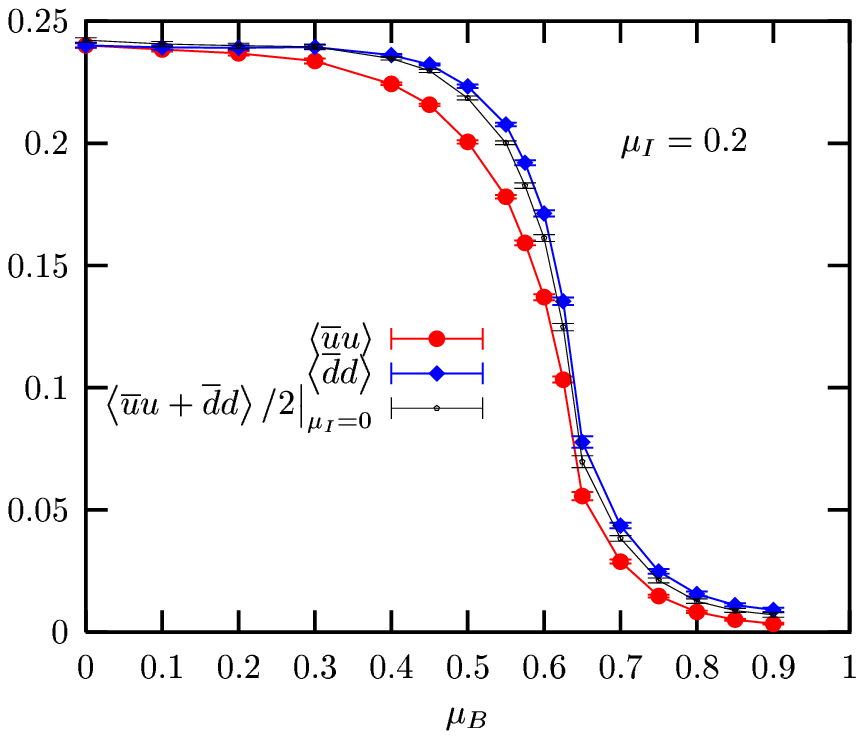}\\
\includegraphics[width=7cm]{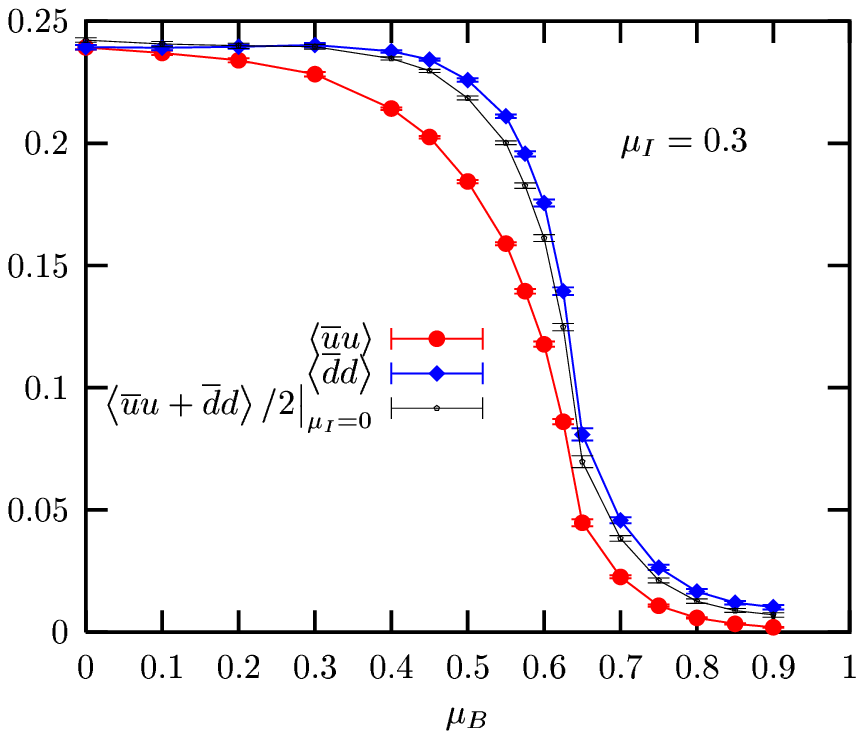}&
\includegraphics[width=7cm]{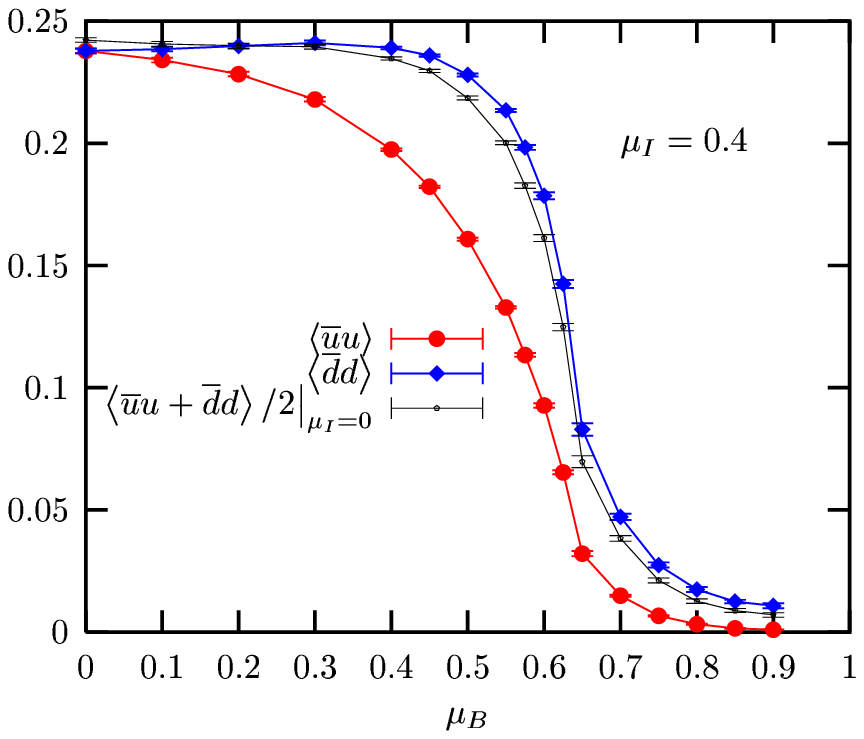}
\end{tabular}
\caption{$\lt<\overline u u\rt>$ and $\lt<\overline d d\rt>$
condensates for various $\mu_B$ and $\mu_I$ on a $12^4$ lattice.}
\label{fig:pbpiso}
\end{figure}
Figure~\ref{fig:pbpiso} illustrates lattice measurements of the up
and down quark condensates 
\be
\lt<\overline u u\rt>,\lt<\overline d d\rt>
\equiv\frac{1}{V}\frac{\partial \ln{
Z}}{\partial m_{u,d}}=\frac{1}{4V}\lt<
{\rm t r}\lt(\ba{c c}&\openone\pm\tau_3\\-\openone\mp\tau_3&\ea\rt){\cal A}^{-1}\rt>
\ee
as functions of $\mu_B$ for various $\mu_I$ measured on a $12^4$
lattice. Although these results are measured on only one 
volume, the speed of these simulations means that is is possible to  
gain fine resolution in $\mu_B$. Consistent with the predictions
of~\cite{Toublan:2003tt}, the two transitions, which are coincident in the
limit that $\mu_I\to0$, separate as $\mu_I$ is increased. This can be
understood by noting that for fixed $\mu_I$ the chemical potential of
the up quark is larger than that of the down, such that as $\mu_B$
increases, $\mu_u$
reaches the critical chemical potential first. It is not clear,
however,  why the
curve of the up condensate deviates from the $\mu_I=0$ solutions more
than that of the down. This effect, not predicted in~\cite{Toublan:2003tt},
could be some finite volume artifact, or a result of the
quenched approximation. As an aside, it has been
argued that the observation of two transitions is an artifact
of the diagonal flavor structure of the NJL model with broken chiral
symmetry and would not be observed in nature. In particular, the
introduction of an instanton-motivated flavor mixing vertex with even
a weak coupling is shown to restore the single
transition~\cite{Frank:2003ve}.

In the diquark sector, we relabel the order parameter $\lt<q q_+\rt>$ 
defined in (\ref{eq:qq}) as $\lt<u d\rt>$, to emphasise the fact that 
condensation occurs between quarks of different flavors. 
\begin{figure}[h]
\centering
\includegraphics[width=12cm]{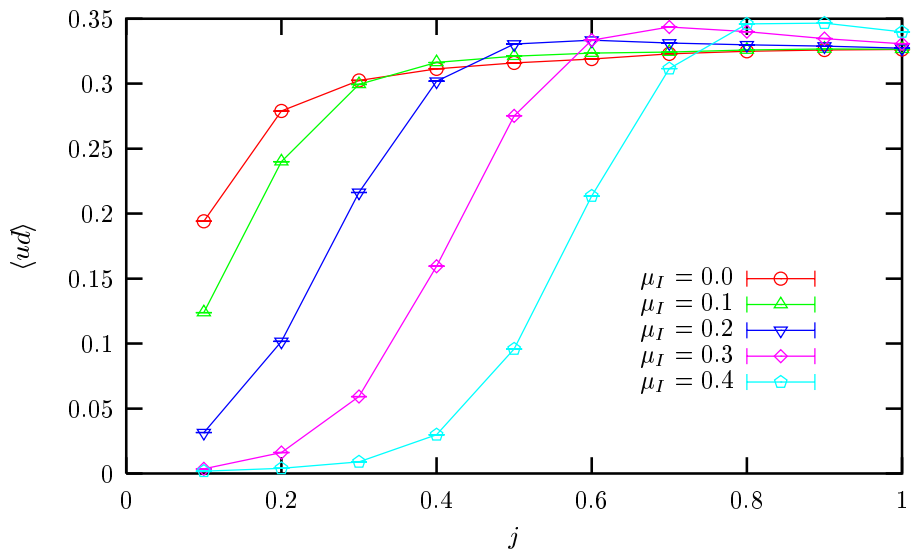}
\caption{$\lt<u d\rt>$ condensate as a function of $j$ for various
$\mu_I$ with $\mu_B=1.0$.}
\label{fig:qqiso}
\end{figure}
Figure~\ref{fig:qqiso} illustrates the $\lt<u d\rt>$ condensate measured
at $\mu_B=1.0$ on $12^4$, $16^4$ and $20^4$ lattices as a function of
$j$ for various values of $\mu_I$. Results are extrapolated to
$T\to0$ as before.  As expected, the effect of significantly
increasing $\mu_I$ for fixed $j$ is to suppress condensation, an
effect that is more pronounced at smaller values of $j$. 
Less straightforward to understand, however, is that at values of $j$ 
above this suppression $\lt<u d(j)\rt>$ appears to increase
slightly with increasing $\mu_I$. By analogy
with~\cite{Bedaque:1999nu}, this could be due to the non-trivial
$O(a)$ color-mixing terms of the staggered quark action, but again
could be due to the crudeness of the 
quenched approximation. 

As with our results at non-zero temperature, because the effect of
increasing the diquark source is to make the superfluid more robust to
setting $\mu_I\neq0$, it is not yet clear how to determine the critical
isospin chemical potential in the limit $j\to0$. 


\section{Summary}
\label{sec:summary}

To our knowledge this is the first systematic non-perturbative study of a
3+1$d$ relativistic system (albeit one with a Lorentz non-invariant cutoff) which
has
a Fermi surface. Our goal has been to study such a system with
phenomenologically reasonable parameters, but with the main focus on 
characterising the high-density ground state.
The principal result is that in the limit $T\to0$ the Fermi
surface is unstable with respect to condensation of diquark pairs in the scalar 
isoscalar channel, leading to a ground state characterised
by a U(1)$_B$-violating order parameter $\langle q q_+\rangle$. 
The resulting energy
gap $\Delta$ which opens up at the Fermi surface is approximately 15\% of the
constituent quark mass scale, in good agreement with self-consistent estimates
made with similar models in continuum approaches. It seems likely that the
model is thus a superfluid described in terms of orthodox BCS theory. 

Detailed quantitative agreement with further aspects of the BCS theory, such as
the prediction (\ref{eq:Tcratio}) for $T_c$, is hard to
verify because of practical difficulties in simultaneously
probing both $T\to0$ and $j\to0$ limits. Finite volume effects in this
system are complicated  to unravel 
because of their separate dependencies on $L_s$ and $L_t$.
As described in Secs.~\ref{sec:finitev}
and \ref{sec:Tneq0}, for the first time we have been able to simulate a
sufficiently broad range of $L_s$, $L_t$ to be able to demonstrate
that the apparent suppression of the order parameter for small $j$ is due 
to our distance from the $L_t\to0$ limit. Note, though, that the ratio $L_t/L_s$
cannot be made arbitrarily large without introducing artifacts due to
discretisation of $k$-space, as shown in Fig.~4 of \cite{Hands:2002mr}.
With the bare parameters used in this study we estimate that volumes
greater than $40^4$ will be needed for quantitative studies of
$T_c$.

Finally, we have made the first exploratory study of a system with non-zero
chemical potentials for both baryon number and isospin, which as described in
Sec.~\ref{sec:mu_Ineq0} is a necessary precondition for an accurate description
of the conditions prevailing inside a neutron star. It is amusing that the sign
problem
comes back to bite, restricting us to considering non-zero isospin density in
the quenched approximation only. Nonetheless, we have been able to observe the
expected suppression of $u d$ pairing at the Fermi surface. Bearing in mind that
phenomenology demands $\mu_I\ll\mu_B$, it is not ruled out that future
simulations could successfully unquench isospin by either reweighting or
analytic continuation in
the small parameter $\mu_I/\mu_B$, much as the conditions in heavy ion
collisions can be accessed by lattice QCD simulations at $\mu_B=0$ 
analytically continued in $\mu_B/T$.


\begin{acknowledgments} 
The simulations presented in sections~\ref{sec:finitev} and
\ref{sec:Tneq0}  were performed
on the Cambridge-Cranfield HPCF SunFire machine.  SJH is supported by
a PPARC Senior Research Fellowship.
\end{acknowledgments}

%
%
\appendix
\section{The Area of the Fermi Surface}
\label{app:area}

In Fig.~\ref{fig:qqEofS} we plot the area of the Fermi surface in
the infinite volume limit as a function of chemical potential
$\mu$. Here we outline the method used to produce this curve. 

In general, one can calculate the area of any surface $S$ by
integrating $\phi=1$ over that surface
\be
\mbox{Surface Area}=\int_{S}1{\rm d}\sigma, 
\ee
where ${\rm d}\sigma$ is an infinitesimal two-dimensional element of
$S$. If the
$z$-component of that surface is single-valued, one may turn this into
a two-dimensional integral in the $x$-$y$ plane
\be
\int_{S}1{\rm d}\sigma\simeq\iint_A\frac{\delta\sigma}{\delta x \delta y}
{\rm d}y{\rm d}y,
\label{eq:1dsig}
\ee
where $A$ is the area projected onto $z=0$ and 
$\delta\sigma$ is the area of a parallelogram tangential to $S$ that projects
onto a square of area $\delta x\delta y$. In the limit that
$\delta\to\rm d$, $\frac{\delta\sigma}{\delta x \delta y}$ is given
by the absolute gradient of $z(x,y)$
\be
\lt|\nabla z(x,y)\rt|
=\lt|\lt(\frac{\partial z}{\partial x},\frac{\partial z}{\partial y},
\frac{\partial z}{\partial z}\rt)\rt|
=\sqrt{\lt(\frac{\partial z}{\partial x}\rt)^2
+\lt(\frac{\partial z}{\partial y}\rt)^2+1}
\ee
and the equality in (\ref{eq:1dsig}) becomes exact.

In calculating the area of the Fermi surface, we consider the section
of momentum space with $0\le k_{x,y,z}\le\frac\pi2$ in which the
surface  is a single-valued function
${k_F}_z({k_F}_x,{k_F}_y)$. From (\ref{eq:Surface}), we find that the
$k_z$-component of the Fermi surface for free fermions at fixed $\mu$ 
is given by
\be
{k_F}_z({k_F}_x,{k_F}_y)=\sin^{-1}\sqrt{C-\sin^2{k_F}_x-\sin^2{k_F}_y}
\label{eq:surfz}
\ee
where the constant $C\equiv\sinh^2\mu-\Sigma^2$ and the value of
$\Sigma(\mu)$ is taken from the large-$N_c$ solution of the gap
equation (\ref{eq:gapeqn}) in the infinite
volume limit. The absolute gradient of this function is
\be
\lt|\nabla {k_F}_z({k_F}_x,{k_F}_y)\rt|=
\sqrt{1+\frac{\cos^2{k_F}_x\sin^2{k_F}_x+\cos^2{k_F}_y\sin^2{k_F}_y}
{(C-\sum_{i=x,y}\sin^2{k_F}_i)(1-C+\sum_{i=x,y}\sin^2{k_F}_i)}}
\ee
and the surface area is given by
\be
\iint_A\lt|\nabla {k_F}_z({k_F}_x,{k_F}_y)\rt|{\rm d}{k_F}_y{\rm d}{k_F}_x,
\ee
which we evaluate numerically. In setting the limits of integration,
we have several cases, determined by the value of $C(\mu)$:  
\renewcommand{\labelenumi}{(\alph{enumi})}
\renewcommand{\labelenumii}{(\roman{enumii})}
\begin{enumerate}
\item
For $C\le0$, (\ref{eq:surfz}) has no real solutions and the surface area is
zero. From the definition of $C$ one can see that this corresponds
physically to the chemical potential being insufficiently large to
allow the existence of a sea of particles with mass $\Sigma$.
\item
For $0<C\le1$, the approximately spherical surface intercepts the
$k_x$ and $k_y$ axes at $\sin^{-1}\sqrt{C}$ and we have one region of
integration. 
\item
For $1<C\le2$, the discretisation of space-time dominates and the
surface no longer intercepts the axes. This is the situation depicted in
 the right-hand 
panel of Fig.~\ref{fig:surface}. (\ref{eq:surfz}) no longer has any
real solutions with $\sin^2k_x+\sin^2k_y\le1$ or $>2$. We must
separate our integral, therefore, into two regions.
\item
For $2<C\le3$, the surface is again approximately spherical, but
with the centre at $(\frac\pi2,\frac\pi2,\frac\pi2)$. 
In this final region, saturation effects are dominant as the surface
area of the sphere decreases with increasing $\mu$ until at $C=3$
(i.e. $\mu\sim\sinh^{-1}\sqrt3\sim1.32$) the area reduces to zero and the
lattice is saturated with fermions.
\end{enumerate}
The limits of integration are listed in table~\ref{tab:arealimits}.
\begin{table}[h]
\begin{center}
\begin{tabular}{|c|c|c|}
\hline\hline
$C$&$k_x$&$k_y$\\
\hline
$\le0$&${\rm Area}=0$&n/a\\
\hline
$0:1$&$0:\sin^{-1}\sqrt C$&$0:\sin^{-1}\sqrt{C-\sin^2k_x}$\\
\hline
$1:2$&$0:\sin^{-1}\sqrt{C-1}$&$\sin^{-1}\sqrt{C-1-\sin^2k_x}:\frac\pi2$\\
&$\sin^{-1}\sqrt{C-1}:\frac\pi2$&$0:\sin^{-1}\sqrt{C-\sin^2k_x}$\\
\hline
$2:3$&$\sin^{-1}\sqrt{C-2}:\frac\pi2$
&$\sin^{-1}\sqrt{C-1-\sin^2k_x}:\frac\pi2$\\
\hline\hline
\end{tabular}
\end{center}
\caption{Limits of integration in calculating the area of the Fermi
surface for different values of $C\equiv\sinh^2\mu-\Sigma^2$.} 
\label{tab:arealimits}
\end{table}

The only limitation of this method is that the
anti-periodicity of the Fermi surface in the region considered implies
that $\nabla {k_F}_z$ diverges at the boundaries $k_z=0$ and
$k_z=\frac\pi2$. Analytically, this divergence is cancelled in the
integral by the infinitesimal size of $\mbox d x$ and $\mbox d y$. In a
numerical solution, however both $\delta x$ and $\delta y$ are
non-vanishing and the integral diverges. 
To overcome this effect we introduce a small ``buffer''
about these boundaries to stop the inclusion of
divergent terms. The size of this buffer is then reduced until its
effect on the solution is negligible. The buffer used to
produce the curve in Fig.~\ref{fig:qqEofS} is $10^{-9}$; once the
curve is evaluated, it is multiplied by an arbitrary constant
(1/45) to allow it to be compared directly with the measured value of 
$\lt<q q_+\rt>$.


\section{The Volume of the Fermi Sea}

In Fig.~\ref{fig:EofS}, the volume of the Fermi sea in the infinite
volume limit is plotted as a function of chemical potential. This
calculation, which is simpler than that for the area of the Fermi
surface, is done by integrating $\phi=1$ numerically over the volume 
bounded by 
(\ref{eq:surfz})
\be
\mbox{Volume}=\iiint_{V}1{\rm d}k_z{\rm d}k_y{\rm d}k_x,
\ee
where again the limits are determined by the value of $C$. These are
listed in table~\ref{tab:volvimits}.

\begin{table}[h]
\begin{center}
\begin{tabular}{|c|c|c|c|}
\hline\hline
$C$&$k_x$&$k_y$&$k_z$\\
\hline
$\le0$&${\rm Volume}=0$&n/a&n/a\\
\hline
$0:1$& $0:\sin^{-1}\sqrt C$& $0:\sin^{-1}\sqrt{C-\sin^2k_x}$
&$0:\sin^{-1}\sqrt{C-\sin^2k_x-\sin^2k_y}$\\
\hline
$1:2$&$0:\sin^{-1}\sqrt{C-1}$&$0:\sin^{-1}\sqrt{C-1-\sin^2k_x}$
&$0:\frac\pi2$\\
&$0:\sin^{-1}\sqrt{C-1}$&$\sin^{-1}\sqrt{C-1-\sin^2k_x}:\frac\pi2$
&$0:\sin^{-1}\sqrt{C-\sin^2k_x-\sin^2k_y}$\\
&$\sin^{-1}\sqrt{C-1}:\frac\pi2$&$0:\sin^{-1}\sqrt{C-\sin^2k_x}$
&$0:\sin^{-1}\sqrt{C-\sin^2k_x-\sin^2k_y}$\\
\hline
$2:3$&$0:\frac\pi2$&$0:\sin^{-1}\sqrt{C-2}$&$0:\frac\pi2$\\
&$0:\sin^{-1}\sqrt{C-2}$&$\sin^{-1}\sqrt{C-2}:\frac\pi2$&$0:\frac\pi2$\\
&$\sin^{-1}\sqrt{C-2}:\frac\pi2$
&$\sin^{-1}\sqrt{C-2}:\sin^{-1}\sqrt{C-1-\sin^2k_x}$&$0:\frac\pi2$\\
&$\sin^{-1}\sqrt{C-2}:\frac\pi2$
&$\sin^{-1}\sqrt{C-1-\sin^2k_x}:\frac\pi2$
&$0:\sin^{-1}\sqrt{C-\sin^2k_x-\sin^2k_y}$\\
\hline\hline
\end{tabular}
\end{center}
\caption{Limits of integration in calculating the volume of the Fermi
sea for different values of $C\equiv\sinh^2\mu-\Sigma^2$.} 
\label{tab:volvimits}
\end{table}

As the integrand, unity, is well behaved over all $k_x$, $k_y$ and
$k_z$, there is no need to introduce a buffer into this calculation.
Once the curve is evaluated, it is normalised such that Volume$(C=3)\equiv1$ 
to allow direct comparison with the large-$N_c$
prediction for $n_B$.


\bibliography{njlgap}

\begin{thebibliography}{49}
\expandafter\ifx\csname natexlab\endcsname\relax\def\natexlab#1{#1}\fi
\expandafter\ifx\csname bibnamefont\endcsname\relax
  \def\bibnamefont#1{#1}\fi
\expandafter\ifx\csname bibfnamefont\endcsname\relax
  \def\bibfnamefont#1{#1}\fi
\expandafter\ifx\csname citenamefont\endcsname\relax
  \def\citenamefont#1{#1}\fi
\expandafter\ifx\csname url\endcsname\relax
  \def\url#1{\texttt{#1}}\fi
\expandafter\ifx\csname urlprefix\endcsname\relax\def\urlprefix{URL }\fi
\providecommand{\bibinfo}[2]{#2}
\providecommand{\eprint}[2][]{\url{#2}}

\bibitem[{\citenamefont{Alford et~al.}(1999)\citenamefont{Alford, Rajagopal,
  and Wilczek}}]{Alford:1998mk}
\bibinfo{author}{\bibfnamefont{M.~G.} \bibnamefont{Alford}},
  \bibinfo{author}{\bibfnamefont{K.}~\bibnamefont{Rajagopal}},
  \bibnamefont{and} \bibinfo{author}{\bibfnamefont{F.}~\bibnamefont{Wilczek}},
  \bibinfo{journal}{Nucl. Phys.} \textbf{\bibinfo{volume}{B537}},
  \bibinfo{pages}{443} (\bibinfo{year}{1999}), \eprint{hep-ph/9804403}.

\bibitem[{\citenamefont{Schafer and Wilczek}(1999)}]{Schafer:1998ef}
\bibinfo{author}{\bibfnamefont{T.}~\bibnamefont{Schafer}} \bibnamefont{and}
  \bibinfo{author}{\bibfnamefont{F.}~\bibnamefont{Wilczek}},
  \bibinfo{journal}{Phys. Rev. Lett.} \textbf{\bibinfo{volume}{82}},
  \bibinfo{pages}{3956} (\bibinfo{year}{1999}), \eprint{hep-ph/9811473}.

\bibitem[{\citenamefont{Rischke}(2004)}]{Rischke:2003mt}
\bibinfo{author}{\bibfnamefont{D.~H.} \bibnamefont{Rischke}},
  \bibinfo{journal}{Prog. Part. Nucl. Phys.} \textbf{\bibinfo{volume}{52}},
  \bibinfo{pages}{197} (\bibinfo{year}{2004}), \eprint{nucl-th/0305030}.

\bibitem[{\citenamefont{Alford et~al.}(2001{\natexlab{a}})\citenamefont{Alford,
  Bowers, and Rajagopal}}]{Alford:2000sx}
\bibinfo{author}{\bibfnamefont{M.~G.} \bibnamefont{Alford}},
  \bibinfo{author}{\bibfnamefont{J.~A.} \bibnamefont{Bowers}},
  \bibnamefont{and}
  \bibinfo{author}{\bibfnamefont{K.}~\bibnamefont{Rajagopal}},
  \bibinfo{journal}{J. Phys.} \textbf{\bibinfo{volume}{G27}},
  \bibinfo{pages}{541} (\bibinfo{year}{2001}{\natexlab{a}}),
  \eprint{hep-ph/0009357}.

\bibitem[{\citenamefont{Rajagopal and Wilczek}(2001)}]{Rajagopal:2000wf}
\bibinfo{author}{\bibfnamefont{K.}~\bibnamefont{Rajagopal}} \bibnamefont{and}
  \bibinfo{author}{\bibfnamefont{F.}~\bibnamefont{Wilczek}},
  \emph{\bibinfo{title}{Handbook of {QCD}}} (\bibinfo{publisher}{World
  Scientific}, \bibinfo{year}{2001}), chap.~\bibinfo{chapter}{35},
  \eprint{hep-ph/0011333}.

\bibitem[{\citenamefont{Alford}(2001)}]{Alford:2001dt}
\bibinfo{author}{\bibfnamefont{M.~G.} \bibnamefont{Alford}},
  \bibinfo{journal}{Ann. Rev. Nucl. Part. Sci.} \textbf{\bibinfo{volume}{51}},
  \bibinfo{pages}{131} (\bibinfo{year}{2001}), \eprint{hep-ph/0102047}.

\bibitem[{\citenamefont{Bowers and Rajagopal}(2002)}]{Bowers:2002xr}
\bibinfo{author}{\bibfnamefont{J.~A.} \bibnamefont{Bowers}} \bibnamefont{and}
  \bibinfo{author}{\bibfnamefont{K.}~\bibnamefont{Rajagopal}},
  \bibinfo{journal}{Phys. Rev.} \textbf{\bibinfo{volume}{D66}},
  \bibinfo{pages}{065002} (\bibinfo{year}{2002}), \eprint{hep-ph/0204079}.

\bibitem[{\citenamefont{Rapp et~al.}(1998)\citenamefont{Rapp, Schafer, Shuryak,
  and Velkovsky}}]{Rapp:1998zu}
\bibinfo{author}{\bibfnamefont{R.}~\bibnamefont{Rapp}},
  \bibinfo{author}{\bibfnamefont{T.}~\bibnamefont{Schafer}},
  \bibinfo{author}{\bibfnamefont{E.~V.} \bibnamefont{Shuryak}},
  \bibnamefont{and}
  \bibinfo{author}{\bibfnamefont{M.}~\bibnamefont{Velkovsky}},
  \bibinfo{journal}{Phys. Rev. Lett.} \textbf{\bibinfo{volume}{81}},
  \bibinfo{pages}{53} (\bibinfo{year}{1998}), \eprint{hep-ph/9711396}.

\bibitem[{\citenamefont{Alford et~al.}(1998)\citenamefont{Alford, Rajagopal,
  and Wilczek}}]{Alford:1998zt}
\bibinfo{author}{\bibfnamefont{M.~G.} \bibnamefont{Alford}},
  \bibinfo{author}{\bibfnamefont{K.}~\bibnamefont{Rajagopal}},
  \bibnamefont{and} \bibinfo{author}{\bibfnamefont{F.}~\bibnamefont{Wilczek}},
  \bibinfo{journal}{Phys. Lett.} \textbf{\bibinfo{volume}{B422}},
  \bibinfo{pages}{247} (\bibinfo{year}{1998}), \eprint{hep-ph/9711395}.

\bibitem[{\citenamefont{Berges and Rajagopal}(1999)}]{Berges:1998rc}
\bibinfo{author}{\bibfnamefont{J.}~\bibnamefont{Berges}} \bibnamefont{and}
  \bibinfo{author}{\bibfnamefont{K.}~\bibnamefont{Rajagopal}},
  \bibinfo{journal}{Nucl. Phys.} \textbf{\bibinfo{volume}{B538}},
  \bibinfo{pages}{215} (\bibinfo{year}{1999}), \eprint{hep-ph/9804233}.

\bibitem[{\citenamefont{Nambu and Jona-Lasinio}(1961)}]{Nambu:1961tp}
\bibinfo{author}{\bibfnamefont{Y.}~\bibnamefont{Nambu}} \bibnamefont{and}
  \bibinfo{author}{\bibfnamefont{G.}~\bibnamefont{Jona-Lasinio}},
  \bibinfo{journal}{Phys. Rev.} \textbf{\bibinfo{volume}{122}},
  \bibinfo{pages}{345} (\bibinfo{year}{1961}).

\bibitem[{\citenamefont{Hands and Kogut}(1998)}]{Hands:1998uf}
\bibinfo{author}{\bibfnamefont{S.}~\bibnamefont{Hands}} \bibnamefont{and}
  \bibinfo{author}{\bibfnamefont{J.~B.} \bibnamefont{Kogut}},
  \bibinfo{journal}{Nucl. Phys.} \textbf{\bibinfo{volume}{B520}},
  \bibinfo{pages}{382} (\bibinfo{year}{1998}), \eprint{hep-lat/9705015}.

\bibitem[{\citenamefont{Hatsuda and Kunihiro}(1985)}]{Hatsuda:1985eb}
\bibinfo{author}{\bibfnamefont{T.}~\bibnamefont{Hatsuda}} \bibnamefont{and}
  \bibinfo{author}{\bibfnamefont{T.}~\bibnamefont{Kunihiro}},
  \bibinfo{journal}{Phys. Rev. Lett.} \textbf{\bibinfo{volume}{55}},
  \bibinfo{pages}{158} (\bibinfo{year}{1985}).

\bibitem[{\citenamefont{Asakawa and Yazaki}(1989)}]{Asakawa:1989bq}
\bibinfo{author}{\bibfnamefont{M.}~\bibnamefont{Asakawa}} \bibnamefont{and}
  \bibinfo{author}{\bibfnamefont{K.}~\bibnamefont{Yazaki}},
  \bibinfo{journal}{Nucl. Phys.} \textbf{\bibinfo{volume}{A504}},
  \bibinfo{pages}{668} (\bibinfo{year}{1989}).

\bibitem[{\citenamefont{Klevansky}(1992)}]{Klevansky:1992qe}
\bibinfo{author}{\bibfnamefont{S.~P.} \bibnamefont{Klevansky}},
  \bibinfo{journal}{Rev. Mod. Phys.} \textbf{\bibinfo{volume}{64}},
  \bibinfo{pages}{649} (\bibinfo{year}{1992}).

\bibitem[{\citenamefont{Bardeen et~al.}(1957)\citenamefont{Bardeen, Cooper, and
  Schrieffer}}]{Bardeen:1957mv}
\bibinfo{author}{\bibfnamefont{J.}~\bibnamefont{Bardeen}},
  \bibinfo{author}{\bibfnamefont{L.~N.} \bibnamefont{Cooper}},
  \bibnamefont{and} \bibinfo{author}{\bibfnamefont{J.~R.}
  \bibnamefont{Schrieffer}}, \bibinfo{journal}{Phys. Rev.}
  \textbf{\bibinfo{volume}{108}}, \bibinfo{pages}{1175} (\bibinfo{year}{1957}).

\bibitem[{\citenamefont{Hands and Morrison}(1999)}]{Hands:1998kk}
\bibinfo{author}{\bibfnamefont{S.~J.} \bibnamefont{Hands}} \bibnamefont{and}
  \bibinfo{author}{\bibfnamefont{S.~E.} \bibnamefont{Morrison}}
  (\bibinfo{collaboration}{UKQCD}), \bibinfo{journal}{Phys. Rev.}
  \textbf{\bibinfo{volume}{D59}}, \bibinfo{pages}{116002}
  (\bibinfo{year}{1999}), \eprint{hep-lat/9807033}.

\bibitem[{\citenamefont{Hands et~al.}(2001{\natexlab{a}})\citenamefont{Hands,
  Lucini, and Morrison}}]{Hands:2000gv}
\bibinfo{author}{\bibfnamefont{S.}~\bibnamefont{Hands}},
  \bibinfo{author}{\bibfnamefont{B.}~\bibnamefont{Lucini}}, \bibnamefont{and}
  \bibinfo{author}{\bibfnamefont{S.}~\bibnamefont{Morrison}},
  \bibinfo{journal}{Phys. Rev. Lett.} \textbf{\bibinfo{volume}{86}},
  \bibinfo{pages}{753} (\bibinfo{year}{2001}{\natexlab{a}}),
  \eprint{hep-lat/0008027}.

\bibitem[{\citenamefont{Hands et~al.}(2002)\citenamefont{Hands, Lucini, and
  Morrison}}]{Hands:2001aq}
\bibinfo{author}{\bibfnamefont{S.~J.} \bibnamefont{Hands}},
  \bibinfo{author}{\bibfnamefont{B.}~\bibnamefont{Lucini}}, \bibnamefont{and}
  \bibinfo{author}{\bibfnamefont{S.~E.} \bibnamefont{Morrison}},
  \bibinfo{journal}{Phys. Rev.} \textbf{\bibinfo{volume}{D65}},
  \bibinfo{pages}{036004} (\bibinfo{year}{2002}), \eprint{hep-lat/0109001}.

\bibitem[{\citenamefont{Hands et~al.}(2003)\citenamefont{Hands, Kogut,
  Strouthos, and Tran}}]{Hands:2003dh}
\bibinfo{author}{\bibfnamefont{S.}~\bibnamefont{Hands}},
  \bibinfo{author}{\bibfnamefont{J.~B.} \bibnamefont{Kogut}},
  \bibinfo{author}{\bibfnamefont{C.~G.} \bibnamefont{Strouthos}},
  \bibnamefont{and} \bibinfo{author}{\bibfnamefont{T.~N.} \bibnamefont{Tran}},
  \bibinfo{journal}{Phys. Rev.} \textbf{\bibinfo{volume}{D68}},
  \bibinfo{pages}{016005} (\bibinfo{year}{2003}), \eprint{hep-lat/0302021}.

\bibitem[{\citenamefont{Nelson}(1983)}]{Nelson}
\bibinfo{author}{\bibfnamefont{D.}~\bibnamefont{Nelson}},
  \bibinfo{journal}{Phase Transitions and Critical Phenomena (eds. Domb and
  Lebowitz)} \textbf{\bibinfo{volume}{7}}, \bibinfo{pages}{1}
  (\bibinfo{year}{1983}).

\bibitem[{\citenamefont{Hands and Walters}(2002)}]{Hands:2002mr}
\bibinfo{author}{\bibfnamefont{S.}~\bibnamefont{Hands}} \bibnamefont{and}
  \bibinfo{author}{\bibfnamefont{D.~N.} \bibnamefont{Walters}},
  \bibinfo{journal}{Phys. Lett.} \textbf{\bibinfo{volume}{B548}},
  \bibinfo{pages}{196} (\bibinfo{year}{2002}), \eprint{hep-lat/0209140}.

\bibitem[{\citenamefont{Nardulli}(2002)}]{Nardulli:2002ma}
\bibinfo{author}{\bibfnamefont{G.}~\bibnamefont{Nardulli}},
  \bibinfo{journal}{Riv. Nuovo Cim.} \textbf{\bibinfo{volume}{25N3}},
  \bibinfo{pages}{1} (\bibinfo{year}{2002}), \eprint{hep-ph/0202037}.

\bibitem[{\citenamefont{Hands et~al.}(2000)}]{Hands:2000ei}
\bibinfo{author}{\bibfnamefont{S.}~\bibnamefont{Hands}} \bibnamefont{et~al.},
  \bibinfo{journal}{Eur. Phys. J.} \textbf{\bibinfo{volume}{C17}},
  \bibinfo{pages}{285} (\bibinfo{year}{2000}), \eprint{hep-lat/0006018}.

\bibitem[{\citenamefont{Karsten and Smit}(1981)}]{Karsten:1981wd}
\bibinfo{author}{\bibfnamefont{L.~H.} \bibnamefont{Karsten}} \bibnamefont{and}
  \bibinfo{author}{\bibfnamefont{J.}~\bibnamefont{Smit}},
  \bibinfo{journal}{Nucl. Phys.} \textbf{\bibinfo{volume}{B183}},
  \bibinfo{pages}{103} (\bibinfo{year}{1981}).

\bibitem[{\citenamefont{Duane et~al.}(1987)\citenamefont{Duane, Kennedy,
  Pendleton, and Roweth}}]{Duane:1987de}
\bibinfo{author}{\bibfnamefont{S.}~\bibnamefont{Duane}},
  \bibinfo{author}{\bibfnamefont{A.~D.} \bibnamefont{Kennedy}},
  \bibinfo{author}{\bibfnamefont{B.~J.} \bibnamefont{Pendleton}},
  \bibnamefont{and} \bibinfo{author}{\bibfnamefont{D.}~\bibnamefont{Roweth}},
  \bibinfo{journal}{Phys. Lett.} \textbf{\bibinfo{volume}{B195}},
  \bibinfo{pages}{216} (\bibinfo{year}{1987}).

\bibitem[{\citenamefont{Gocksch}(1988)}]{Gocksch:1988ha}
\bibinfo{author}{\bibfnamefont{A.}~\bibnamefont{Gocksch}},
  \bibinfo{journal}{Phys. Rev.} \textbf{\bibinfo{volume}{D37}},
  \bibinfo{pages}{1014} (\bibinfo{year}{1988}).

\bibitem[{\citenamefont{Stephanov}(1996)}]{Stephanov:1996ki}
\bibinfo{author}{\bibfnamefont{M.~A.} \bibnamefont{Stephanov}},
  \bibinfo{journal}{Phys. Rev. Lett.} \textbf{\bibinfo{volume}{76}},
  \bibinfo{pages}{4472} (\bibinfo{year}{1996}), \eprint{hep-lat/9604003}.

\bibitem[{\citenamefont{Barbour et~al.}(1999)\citenamefont{Barbour, Hands,
  Kogut, Lombardo, and Morrison}}]{Barbour:1999mc}
\bibinfo{author}{\bibfnamefont{I.}~\bibnamefont{Barbour}},
  \bibinfo{author}{\bibfnamefont{S.}~\bibnamefont{Hands}},
  \bibinfo{author}{\bibfnamefont{J.~B.} \bibnamefont{Kogut}},
  \bibinfo{author}{\bibfnamefont{M.-P.} \bibnamefont{Lombardo}},
  \bibnamefont{and} \bibinfo{author}{\bibfnamefont{S.}~\bibnamefont{Morrison}},
  \bibinfo{journal}{Nucl. Phys.} \textbf{\bibinfo{volume}{B557}},
  \bibinfo{pages}{327} (\bibinfo{year}{1999}), \eprint{hep-lat/9902033}.

\bibitem[{\citenamefont{Glendenning}(1992)}]{Glendenning:1992vb}
\bibinfo{author}{\bibfnamefont{N.~K.} \bibnamefont{Glendenning}},
  \bibinfo{journal}{Phys. Rev.} \textbf{\bibinfo{volume}{D46}},
  \bibinfo{pages}{1274} (\bibinfo{year}{1992}).

\bibitem[{\citenamefont{Kogut and Strouthos}(2001)}]{Kogut:1999um}
\bibinfo{author}{\bibfnamefont{J.~B.} \bibnamefont{Kogut}} \bibnamefont{and}
  \bibinfo{author}{\bibfnamefont{C.~G.} \bibnamefont{Strouthos}},
  \bibinfo{journal}{Phys. Rev.} \textbf{\bibinfo{volume}{D63}},
  \bibinfo{pages}{054502} (\bibinfo{year}{2001}), \eprint{hep-lat/9904008}.

\bibitem[{\citenamefont{Aloisio et~al.}(2001)\citenamefont{Aloisio, Azcoiti,
  Di~Carlo, Galante, and Grillo}}]{Aloisio:2000rb}
\bibinfo{author}{\bibfnamefont{R.}~\bibnamefont{Aloisio}},
  \bibinfo{author}{\bibfnamefont{V.}~\bibnamefont{Azcoiti}},
  \bibinfo{author}{\bibfnamefont{G.}~\bibnamefont{Di~Carlo}},
  \bibinfo{author}{\bibfnamefont{A.}~\bibnamefont{Galante}}, \bibnamefont{and}
  \bibinfo{author}{\bibfnamefont{A.~F.} \bibnamefont{Grillo}},
  \bibinfo{journal}{Nucl. Phys.} \textbf{\bibinfo{volume}{B606}},
  \bibinfo{pages}{322} (\bibinfo{year}{2001}), \eprint{hep-lat/0011079}.

\bibitem[{\citenamefont{Kogut et~al.}(2003)\citenamefont{Kogut, Toublan, and
  Sinclair}}]{Kogut:2003ju}
\bibinfo{author}{\bibfnamefont{J.~B.} \bibnamefont{Kogut}},
  \bibinfo{author}{\bibfnamefont{D.}~\bibnamefont{Toublan}}, \bibnamefont{and}
  \bibinfo{author}{\bibfnamefont{D.~K.} \bibnamefont{Sinclair}},
  \bibinfo{journal}{Phys. Rev.} \textbf{\bibinfo{volume}{D68}},
  \bibinfo{pages}{054507} (\bibinfo{year}{2003}), \eprint{hep-lat/0305003}.

\bibitem[{\citenamefont{Hands et~al.}(2001{\natexlab{b}})\citenamefont{Hands,
  Montvay, Scorzato, and Skullerud}}]{Hands:2001ee}
\bibinfo{author}{\bibfnamefont{S.~J.} \bibnamefont{Hands}},
  \bibinfo{author}{\bibfnamefont{I.}~\bibnamefont{Montvay}},
  \bibinfo{author}{\bibfnamefont{L.}~\bibnamefont{Scorzato}}, \bibnamefont{and}
  \bibinfo{author}{\bibfnamefont{J.}~\bibnamefont{Skullerud}},
  \bibinfo{journal}{Eur. Phys. J.} \textbf{\bibinfo{volume}{C22}},
  \bibinfo{pages}{451} (\bibinfo{year}{2001}{\natexlab{b}}),
  \eprint{hep-lat/0109029}.

\bibitem[{\citenamefont{Schmitt et~al.}(2002)\citenamefont{Schmitt, Wang, and
  Rischke}}]{Schmitt:2002sc}
\bibinfo{author}{\bibfnamefont{A.}~\bibnamefont{Schmitt}},
  \bibinfo{author}{\bibfnamefont{Q.}~\bibnamefont{Wang}}, \bibnamefont{and}
  \bibinfo{author}{\bibfnamefont{D.~H.} \bibnamefont{Rischke}},
  \bibinfo{journal}{Phys. Rev.} \textbf{\bibinfo{volume}{D66}},
  \bibinfo{pages}{114010} (\bibinfo{year}{2002}), \eprint{nucl-th/0209050}.

\bibitem[{\citenamefont{Elitzur}(1975)}]{Elitzur:1975im}
\bibinfo{author}{\bibfnamefont{S.}~\bibnamefont{Elitzur}},
  \bibinfo{journal}{Phys. Rev.} \textbf{\bibinfo{volume}{D12}},
  \bibinfo{pages}{3978} (\bibinfo{year}{1975}).

\bibitem[{\citenamefont{Boyd et~al.}(1992)\citenamefont{Boyd, Karsch, and
  Gupta}}]{Boyd:1992uk}
\bibinfo{author}{\bibfnamefont{G.}~\bibnamefont{Boyd}},
  \bibinfo{author}{\bibfnamefont{F.}~\bibnamefont{Karsch}}, \bibnamefont{and}
  \bibinfo{author}{\bibfnamefont{S.}~\bibnamefont{Gupta}},
  \bibinfo{journal}{Nucl. Phys.} \textbf{\bibinfo{volume}{B385}},
  \bibinfo{pages}{481} (\bibinfo{year}{1992}).

\bibitem[{\citenamefont{Amore et~al.}(2002)\citenamefont{Amore, Birse,
  McGovern, and Walet}}]{Amore:2001uf}
\bibinfo{author}{\bibfnamefont{P.}~\bibnamefont{Amore}},
  \bibinfo{author}{\bibfnamefont{M.~C.} \bibnamefont{Birse}},
  \bibinfo{author}{\bibfnamefont{J.~A.} \bibnamefont{McGovern}},
  \bibnamefont{and} \bibinfo{author}{\bibfnamefont{N.~R.} \bibnamefont{Walet}},
  \bibinfo{journal}{Phys. Rev.} \textbf{\bibinfo{volume}{D65}},
  \bibinfo{pages}{074005} (\bibinfo{year}{2002}), \eprint{hep-ph/0110267}.

\bibitem[{\citenamefont{Walters and Hands}(2004)}]{Walters:2003it}
\bibinfo{author}{\bibfnamefont{D.~N.} \bibnamefont{Walters}} \bibnamefont{and}
  \bibinfo{author}{\bibfnamefont{S.}~\bibnamefont{Hands}}
  (\bibinfo{year}{2004}), \eprint{hep-lat/0308030}.

\bibitem[{\citenamefont{Walters}(2003)}]{Walters:2003nn}
\bibinfo{author}{\bibfnamefont{D.~N.} \bibnamefont{Walters}}
  (\bibinfo{year}{2003}), \eprint{hep-lat/0310038}.

\bibitem[{\citenamefont{Alford et~al.}(2001{\natexlab{b}})\citenamefont{Alford,
  Bowers, and Rajagopal}}]{Alford:2000ze}
\bibinfo{author}{\bibfnamefont{M.~G.} \bibnamefont{Alford}},
  \bibinfo{author}{\bibfnamefont{J.~A.} \bibnamefont{Bowers}},
  \bibnamefont{and}
  \bibinfo{author}{\bibfnamefont{K.}~\bibnamefont{Rajagopal}},
  \bibinfo{journal}{Phys. Rev.} \textbf{\bibinfo{volume}{D63}},
  \bibinfo{pages}{074016} (\bibinfo{year}{2001}{\natexlab{b}}),
  \eprint{hep-ph/0008208}.

\bibitem[{\citenamefont{Alford et~al.}(2000)\citenamefont{Alford, Berges, and
  Rajagopal}}]{Alford:1999pb}
\bibinfo{author}{\bibfnamefont{M.~G.} \bibnamefont{Alford}},
  \bibinfo{author}{\bibfnamefont{J.}~\bibnamefont{Berges}}, \bibnamefont{and}
  \bibinfo{author}{\bibfnamefont{K.}~\bibnamefont{Rajagopal}},
  \bibinfo{journal}{Nucl. Phys.} \textbf{\bibinfo{volume}{B571}},
  \bibinfo{pages}{269} (\bibinfo{year}{2000}), \eprint{hep-ph/9910254}.

\bibitem[{\citenamefont{Bedaque}(2002)}]{Bedaque:1999nu}
\bibinfo{author}{\bibfnamefont{P.~F.} \bibnamefont{Bedaque}},
  \bibinfo{journal}{Nucl. Phys.} \textbf{\bibinfo{volume}{A697}},
  \bibinfo{pages}{569} (\bibinfo{year}{2002}), \eprint{hep-ph/9910247}.

\bibitem[{\citenamefont{Clogsten}(1962)}]{Clogsten}
\bibinfo{author}{\bibfnamefont{A.}~\bibnamefont{Clogsten}},
  \bibinfo{journal}{Phys. Rev. Lett.} \textbf{\bibinfo{volume}{9}},
  \bibinfo{pages}{266} (\bibinfo{year}{1962}).

\bibitem[{\citenamefont{Chandrasekhar}(1962)}]{Chandrasekhar}
\bibinfo{author}{\bibfnamefont{B.}~\bibnamefont{Chandrasekhar}},
  \bibinfo{journal}{App. Phys. Lett.} \textbf{\bibinfo{volume}{1}},
  \bibinfo{pages}{7} (\bibinfo{year}{1962}).

\bibitem[{\citenamefont{Larkin and Ovchinnikiv}(1964)}]{Larkin}
\bibinfo{author}{\bibfnamefont{A.}~\bibnamefont{Larkin}} \bibnamefont{and}
  \bibinfo{author}{\bibfnamefont{Y.}~\bibnamefont{Ovchinnikiv}},
  \bibinfo{journal}{Zh. Eksp. Teor. Fiz.} \textbf{\bibinfo{volume}{47}},
  \bibinfo{pages}{1136} (\bibinfo{year}{1964}).

\bibitem[{\citenamefont{Fulde and Ferrel}(1964)}]{Fulde}
\bibinfo{author}{\bibfnamefont{P.}~\bibnamefont{Fulde}} \bibnamefont{and}
  \bibinfo{author}{\bibfnamefont{R.}~\bibnamefont{Ferrel}},
  \bibinfo{journal}{Phys. Rev.} \textbf{\bibinfo{volume}{135}},
  \bibinfo{pages}{A550} (\bibinfo{year}{1964}).

\bibitem[{\citenamefont{Toublan and Kogut}(2003)}]{Toublan:2003tt}
\bibinfo{author}{\bibfnamefont{D.}~\bibnamefont{Toublan}} \bibnamefont{and}
  \bibinfo{author}{\bibfnamefont{J.~B.} \bibnamefont{Kogut}},
  \bibinfo{journal}{Phys. Lett.} \textbf{\bibinfo{volume}{B564}},
  \bibinfo{pages}{212} (\bibinfo{year}{2003}), \eprint{hep-ph/0301183}.

\bibitem[{\citenamefont{Frank et~al.}(2003)\citenamefont{Frank, Buballa, and
  Oertel}}]{Frank:2003ve}
\bibinfo{author}{\bibfnamefont{M.}~\bibnamefont{Frank}},
  \bibinfo{author}{\bibfnamefont{M.}~\bibnamefont{Buballa}}, \bibnamefont{and}
  \bibinfo{author}{\bibfnamefont{M.}~\bibnamefont{Oertel}},
  \bibinfo{journal}{Phys. Lett.} \textbf{\bibinfo{volume}{B562}},
  \bibinfo{pages}{221} (\bibinfo{year}{2003}), \eprint{hep-ph/0303109}.

\end{thebibliography}

\end{document}